\documentclass[manuscript=letter]{achemso}

\usepackage[T1]{fontenc}
\usepackage{siunitx}
\usepackage{hyperref,doi}
\usepackage{pdfpages}

\hypersetup{colorlinks=true,citecolor=blue,linkcolor=blue,urlcolor=blue}
\setkeys{acs}{doi = true}

\newcommand{\MoTe}{MoTe$_{2}$~}
\newcommand{\uJcm}{\si{\micro\joule\per\centi\metre\squared}~}
\newcommand{\cm}{\si{\per\centi\metre}~}
\newcommand{\um}{\si{\micro\metre}~}
\newcommand{\Ag}{$A_{\mathrm{1g}}$~}
\newcommand{\dR}{$\Delta R/R$~}

\author{Charles J. Sayers}
\affiliation[Politecnico di Milano]
{Dipartimento di Fisica, Politecnico di Milano, 20133 Milano, Italy}
\email{charles.sayers@polimi.it}

\author{Armando Genco}
\affiliation[Politecnico di Milano]
{Dipartimento di Fisica, Politecnico di Milano, 20133 Milano, Italy}

\author{Chiara Trovatello}
\affiliation[Politecnico di Milano]
{Dipartimento di Fisica, Politecnico di Milano, 20133 Milano, Italy}
\altaffiliation{Department of Mechanical Engineering, Columbia University, New York, New York 10027, United States}

\author{Stefano Dal Conte}
\affiliation[Politecnico di Milano]
{Dipartimento di Fisica, Politecnico di Milano, 20133 Milano, Italy}

\author{Vladislav Khaustov}
\affiliation[IIT]
{Center for Nanotechnology Innovation @ NEST, Istituto Italiano di Tecnologia, 56127 Pisa, Italy}
\altaffiliation{Scuola Normale Superiore, Piazza San Silvestro 12, 56127 Pisa, Italy}

\author{Jorge Cervantes-Villanueva}
\affiliation[University of Valencia]
{Institute of Materials Science (ICMUV), University of Valencia, Catedrático Beltrán 2, E-46980 Valencia, Spain}

\author{Davide Sangalli}
\affiliation[University of Valencia]
{Istituto di Struttura della Materia-CNR (ISM-CNR), Division of Ultrafast Processes in Materials (FLASHit), Area della Ricerca di Roma 1, Monterotondo Scalo, Italy}

\author{Alejandro Molina-Sanchez}
\affiliation[University of Valencia]
{Institute of Materials Science (ICMUV), University of Valencia, Catedrático Beltrán 2, E-46980 Valencia, Spain}

\author{Camilla Coletti}
\affiliation[IIT]
{Center for Nanotechnology Innovation @ NEST, Istituto Italiano di Tecnologia, 56127 Pisa, Italy}
\altaffiliation{Graphene Labs, Istituto Italiano di Tecnologia, 16163 Genova, Italy}

\author{Christoph Gadermaier}
\affiliation[Politecnico di Milano]
{Dipartimento di Fisica, Politecnico di Milano, 20133 Milano, Italy}

\author{Giulio Cerullo}
\affiliation[Politecnico di Milano]
{Dipartimento di Fisica, Politecnico di Milano, 20133 Milano, Italy}

\title{Strong Coupling of Coherent Phonons to Excitons in Semiconducting Monolayer MoTe$_2$.}

\begin{document}

\newpage



\begin{abstract}
The coupling of the electron system to lattice vibrations and their time-dependent control and detection provides unique insight into the non-equilibrium physics of semiconductors. Here, we investigate the ultrafast transient response of semiconducting monolayer 2$H$-\MoTe encapsulated with $h$BN using broadband optical pump-probe microscopy. The sub-40-fs pump pulse triggers extremely intense and long-lived coherent oscillations in the spectral region of the A’ and B’ exciton resonances, up to $\sim$20\% of the maximum transient signal, due to the displacive excitation of the out-of-plane \Ag phonon. \textit{Ab initio} calculations reveal a dramatic rearrangement of the optical absorption of monolayer \MoTe induced by an out-of-plane stretching and compression of the crystal lattice, consistent with an \Ag-type oscillation. Our results highlight the extreme sensitivity of the optical properties of monolayer TMDs to small structural modifications and their manipulation with light.
\end{abstract}


\section{Introduction}

Electron-phonon coupling is crucially important to many phenomena in condensed matter, such as carrier scattering in transport\cite{Bardeen1950,Fischetti1996,Gunst2016}, the relaxation of photoinduced non-equilibrium quasiparticle populations\cite{Allen1987,Gadermaier2010,He2020,Molina-Sanchez2017,Selig2016}, and electronic order emerging at low temperatures\cite{Porer2014,Hedayat2019,Maklar2021,Sayers2022}. Photoexcitation of coherent phonons (CPs) using ultrashort light pulses enables fundamental insight into electron-phonon coupling via their excitation and detection mechanisms\cite{Cerullo2007,Dekorsy2010,Ishioka2010}, which has led to fundamental discoveries such as elucidating the role of vibrational coherence in the primary event of vision\cite{Wang2011}, detecting coherent Bloch oscillations in coupled semiconductor quantum wells\cite{Feldmann1992}, or demonstrating THz radiation emission due to the macroscopic polarization originating from CPs\cite{Dekorsy1995}. There is potential for applications in sensors, actuators, and transducers\cite{Li2013,Lanzillotti-Kimura2018,Baldini2018,Baldini2019} operating at frequencies up to several THz\cite{Hase2013}. In semiconductors with large exciton binding energies, photoexcitation of CPs has provided vital information about exciton-phonon coupling\cite{Luer2009,Ni2017,Trovatello2020a,Mor2021}.

Two-dimensional semiconductors such as monolayer transition metal dichalcogenides (TMDs) combine strong light-matter interaction and multifaceted exciton and valley physics with a great potential for applications in energy harvesting and information processing\cite{Radisavljevic2011,Wang2012,Fiori2014,Jariwala2014,Mak2016,Koppens2014}. 
An important tool in the investigation of exciton-coherent phonon coupling in these materials is femtosecond transient absorption (TA) spectroscopy. Here, an initial ultrashort pump pulse photoexcites the sample, while a second delayed probe pulse is used to measure the transient change in the optical response. This allows for tracking the temporal evolution of the non-equilibrium quasiparticle populations, which manifests as the characteristic decay time of the TA signal. In  WSe$_2$, it has previously been shown that coherent oscillations of the out-of-plane A$_\mathrm{1g}$ phonon mode introduce a small modulation ($\approx10^{-3}$ of the transient signal) on the electronic relaxation at the optical bandgap\cite{Jeong2016}. Similarly in monolayer MoS$_2$, the oscillatory modulation of the TA signal has also been ascribed to CPs belonging to the \Ag mode, which exhibit a relatively small amplitude across the spectral region of the A and B exciton resonances\cite{Jeong2016}, but a significant modulation of approximately 2\% of the signal maximum around the C exciton\cite{Trovatello2020a}.

\MoTe is a TMD with two well-known thermodynamically stable polymorphs with distinct electronic properties. Its semimetallic phase\cite{Cho2015,Keum2015}, (1$T$’ above and $T_d$ below 240 K) exhibits large magnetoresistance\cite{Lee2018}, ferroelectricity\cite{Jindal2023} and superconductivity\cite{Qi2016}. The semiconducting 2$H$ phase, on the other hand, has an indirect bandgap in the bulk which becomes direct ($\sim$1.1 eV) towards the monolayer limit \cite{Ruppert2014,Lezama2015}, thus expanding the potential functionality of TMDs into the near-infrared. Furthermore, a high carrier mobility\cite{Mleczko2019}, strong spin-orbit coupling\cite{Pradhan2014}, valley selectivity\cite{Li2018}, and ambipolar transistor behavior\cite{Lin2014}, make it a promising candidate for near-infrared optoelectronics, photovoltaics, and unconventional information encoding such as spin- or valleytronics. However, since its lower chemical stability has been overcome only recently by encapsulation using few-layer $h$BN\cite{Pace2021}, \MoTe is significantly less studied than its sulfur and selenium analogues. Previous TA experiments on 2$H$-\MoTe ranging from 5 layers to bulk, with various probe energy ranges from 1.0 to 2.6 eV, have elucidated the dynamics of several excitonic transitions, but without detecting any CP signature thus far\cite{Chi2019,Perlangeli2020,Schulzetenberg2021}. Optical pump-core level (XUV) probe spectroscopy, on the other hand, has revealed a strong oscillatory signal contribution dominated by the out-of-plane \Ag mode with a smaller $E_{1g}$ component\cite{Attar2020}, where the greatest oscillation amplitude was observed for transitions from the Te-4$d_{5/2}$ levels to the conduction band.

Here, we employ broadband TA microscopy to study the coupling of the out-of-plane \Ag vibrational mode in semiconducting monolayer 2$H$-\MoTe to several excitonic resonances. Our work is supported by \textit{ab initio} simulations which combine density functional and perturbation theory (DFT/DFPT) with many-body perturbation theory (GW+BSE). We find an exceptionally strong and long-lived oscillatory signal contribution, which is rarely observed in semiconducting TMDs. Our broadband probe combined with an excellent temporal resolution of ($\leq$ 40 fs) reveals the spectral dependence of the amplitude and phase of CPs with exceptional clarity. Our simulations confirm a strong modulation of the electronic band structure and, consequently, the absorption spectrum by out-of-plane atomic motion, allowing the theoretical prediction of the spectral profile of the oscillations in excellent agreement with the experimental observations. Our results demonstrate how the optical properties of monolayer \MoTe in the visible and near-infrared range are highly susceptible to manipulation via small structural modifications, and how these can be controlled optically using ultrashort light pulses.


\section{Results and discussion}

\begin{figure}
    \includegraphics[width=0.5\linewidth]{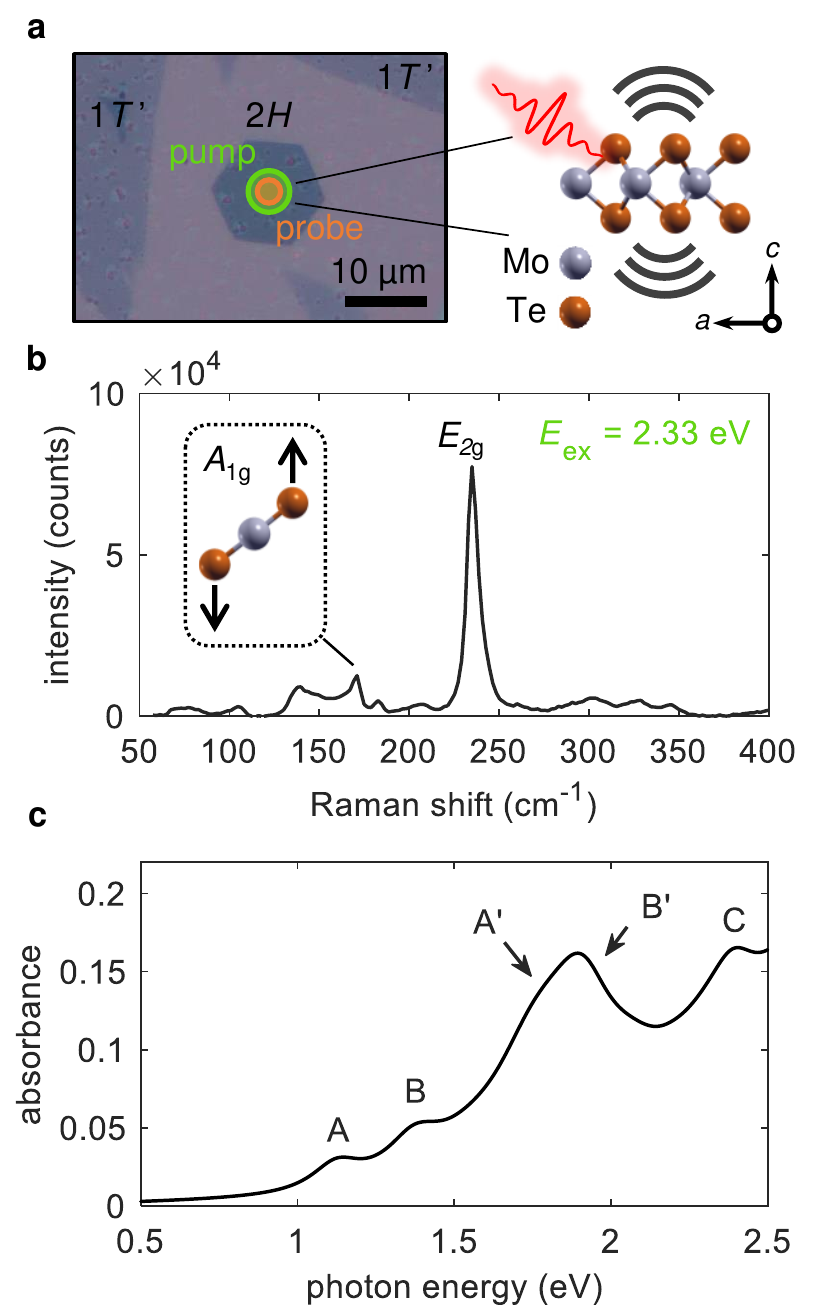}
    \caption{\textbf{Optical properties of monolayer 2$H$-MoTe$_{2}$.} (\textbf{a}) Microscope image of $h$BN encapsulated monolayer MoTe$_{2}$ samples on Si/SiO$_2$ (left). Optical pump-probe microscopy experiments were performed on the 2$H$ region with pump ($\sim$ 5 \um) and probe ($\sim$ 3 \um) beam diameters, as illustrated. The optical pulse launches an intense out-of-plane ($c$-axis) vibration of the lattice (right). (\textbf{b}) Raman spectrum of the \MoTe sample measured with 532 nm ($\sim$ 2.33 eV) excitation. The out-of-plane vibration with \Ag symmetry is highlighted. (\textbf{c}) Optical absorption spectrum for the equilibrium structure from \textit{ab initio} calculations. Optical transitions are labelled according to the convention of Ref.\cite{Wilson1969}.}
  \label{fig:intro}
\end{figure}

Samples of \MoTe were synthesized by chemical vapour deposition on Si/SiO$_2$ and encapsulated with few-layer $h$BN according to the methods in Ref~\cite{Pace2021}. The procedure yields flakes of both polymorphs, which are easily distinguished by their shape; elongated for 1$T'$, or hexagonal for 2$H$. Since the 2$H$ flakes have a lateral size of only a few microns, we employed a broadband optical pump-probe microscope\cite{Wang2021} whereby pump and probe are focused onto the sample using an objective lens, as illustrated in Figure~\ref{fig:intro}a, providing a spatial resolution of $\sim$3 \um. The Raman spectrum measured on the same flake, shown in Figure~\ref{fig:intro}b, confirms the semiconducting 2$H$ polymorph, with the most prominent peak at 235 \cm originating from the in-plane $E_{\mathrm{2g}}$ phonon. A further peak, seen here at 170 \cm, is associated with the out-of-plane \Ag phonon and has been shown in previous studies to be clearly visible for excitation at 633 nm (1.96 eV), but much weaker for excitation at 532 nm (2.33 eV)\cite{Ruppert2014}. The absence of modes at 120 and 290 \cm confirms the flake to be monolayer\cite{Ruppert2014,Froehlicher2015}. The calculated absorption spectrum of monolayer 2$H$-\MoTe, shown in Figure~\ref{fig:intro}c, exhibits a series of excitonic resonances, whose energies and relative intensities match closely with previously measured spectra\cite{Wilson1969,Ruppert2014}.

We now investigate the transient dynamics of monolayer \MoTe at $T$ = 10 K after photoexcitation with a pump centered around $\sim$2.36 eV. We measure the differential reflectance, \dR with a broadband probe in the range $\sim$ 1.7 - 2.6 eV at variable delay after excitation. The pump and probe beams are focused and spatially overlapped on the sample, as indicated in Figure~\ref{fig:intro}a. The probe is detected in backscattering geometry after interaction with the $h$BN-MoTe$_2$-Si/SiO$_2$ sample stack. The dominant effect of photoexcitation is a change in the absorption spectrum of the \MoTe layer, and hence we assume the measured differential reflectance, \dR to be proportional to -$\Delta A$, i.e. the negative change in absorbance of \MoTe.

\begin{figure}
    \includegraphics[width=1.0\linewidth]{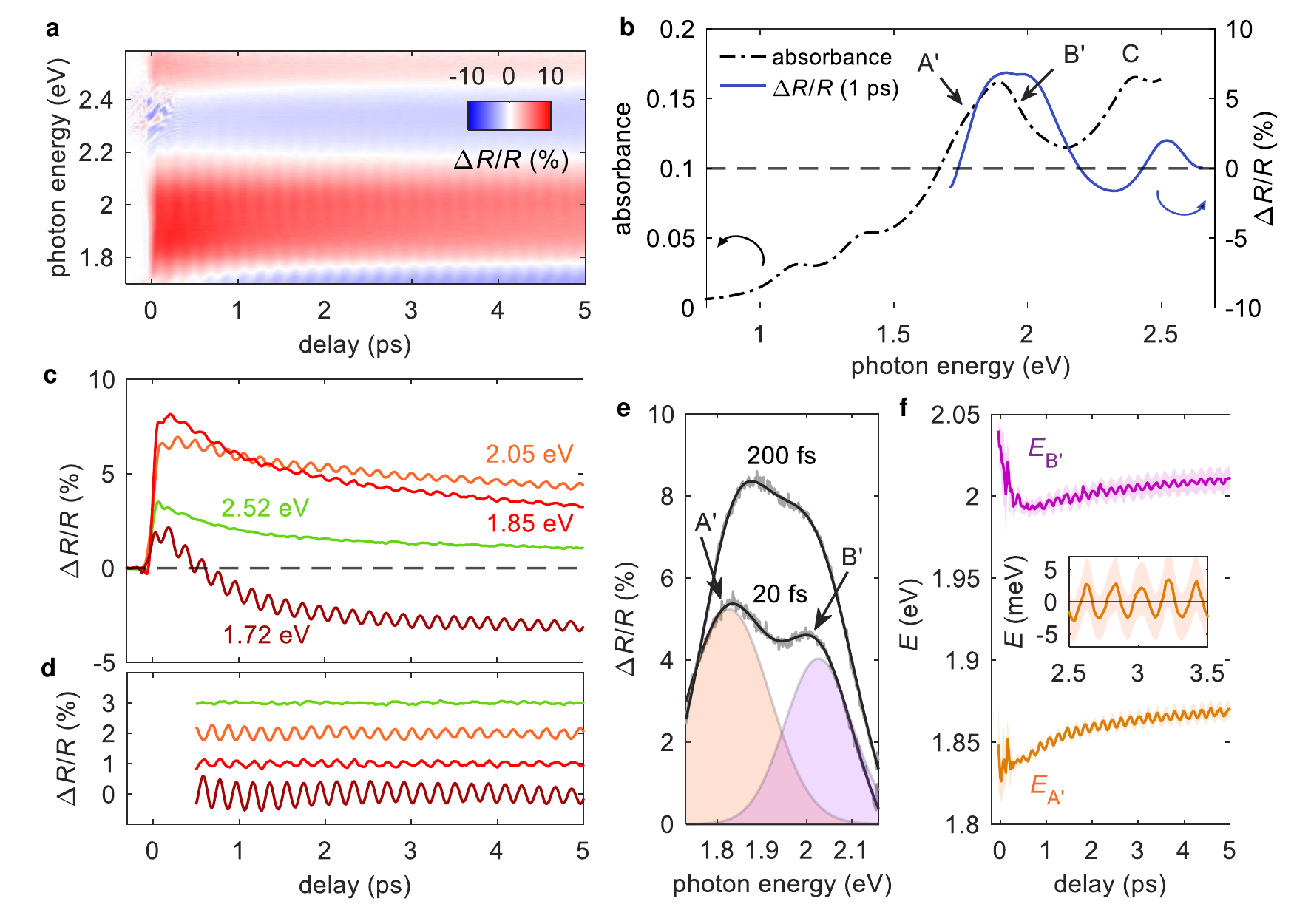}
    \caption{\textbf{Transient optical response of monolayer 2$H$-MoTe$_{2}$.} (\textbf{a}) Broadband differential reflectance (\dR) maps following excitation with a pump photon energy of 2.36 eV and fluence of 500 \uJcm. The sample temperature was 10 K. (\textbf{b}) Transient \dR spectrum (right axis) at 1 ps delay compared to the calculated optical absorbance of the equilibrium structure (left axis). (\textbf{c}) Dynamics extracted at various probe photon energies, as indicated. (\textbf{d}) Isolated coherent component of the \dR signal after subtraction of a bi-exponential fit to the incoherent dynamics in panel c. Data are offset for clarity. (\textbf{e}) \dR spectra at early times (20 and 200 fs) showing two positive peaks related to the A' and B' transitions. The bold solid lines are fits to the data using a multiple Gaussian procedure, where the shaded areas show the two individual components for the A' (orange) and B' (purple) peaks. (\textbf{f}) Temporal evolution of the peak centre energies, $E_{\mathrm{A'}}$ and $E_{\mathrm{B'}}$ obtained from the fitting in panel e. The inset shows the energy modulation of the $E_{\mathrm{A'}}$ peak after subtraction of a bi-exponential. The shaded areas are the associated fitting errors.}
  \label{fig:dynamics}
\end{figure}

The transient \dR spectra, shown over the first 5 ps in Figure~\ref{fig:dynamics}a, exhibit two positive bands of increased reflectivity upon photoexcitation and two negative bands of decreased reflectivity. The positive \dR signal is ascribed to photobleaching (PB) of the excitonic transitions. The PB peaks from 1.75 to 2.2 eV and above 2.4 eV match the positions of the A’, B’ and C excitonic resonances according to Ref\cite{Ruppert2014} (see also Figure~\ref{fig:dynamics}b), suggesting a reduced absorption due to Pauli blocking. The negative \dR signal instead originates from exciton energy renormalization, which causes a shift of the transition and induces a change of sign in the pump-probe signal, from 2.2 to 2.4 eV and below 1.75 eV \cite{Pogna2016,Trovatello2022}. Immediately after photoexcitation an increase of the electronic temperature broadens the exciton lineshape, which subsequently narrows at longer delay times. The extracted exciton dynamics are shown in Figure.~\ref{fig:dynamics}c. After a nearly instantaneous rise time, i.e. limited by the temporal resolution of the experiment, the PB signal related to the A' and B' resonances decays with time constants of the order $\tau_{fast}$ = (0.7 - 0.8) ps and $\tau_{slow}$ = (6.0 - 8.0) ps (see Supporting Information, Figure~S1), without further significant changes of spectral shape.

We now turn our attention to the oscillations, clearly visible in the spectral region from 1.7 to 2.3 eV (see Figure~\ref{fig:dynamics}a,c), which we assign to photoexcited CPs\cite{Trovatello2020a}. Such oscillations have not been observed in previous work on multilayer \MoTe for a probe in the visible range\cite{Chi2019,Perlangeli2020,Schulzetenberg2021}. Here, however, we find an exceptionally strong coherent (oscillatory) component with a magnitude of up to 20\% of the maximum \dR signal, which is an order of magnitude greater than the CP amplitude (2\%) found in monolayer MoS$_2$\cite{Trovatello2020a}. The isolated oscillatory component, shown in Figure~\ref{fig:dynamics}d, obtained by subtracting the incoherent (non-oscillatory) signal contribution with a bi-exponential decay, can be fitted with a single damped cosine term (see Supporting Information, Figure~S3). This oscillatory component exhibits an unusually long damping time of $\tau_{damp}$ = (6.25 $\pm$ 0.25) ps, suggesting that vibrational dephasing via defects is weak\cite{Hase2000}. The CP mode, measured at 10 K, has a period of 194 fs which corresponds to a frequency of 5.15 THz ($\sim$ 172 \cm), and an energy of 21 meV. The \Ag Raman mode frequency in Figure~\ref{fig:intro}b, measured instead at 295 K, is 5.10 THz ($\sim$ 170 \cm), consistent with a temperature-induced mode softening. Therefore, based on this agreement, and previous observations in pump-probe studies of TMDs\cite{Trovatello2020a,Sayers2022}, we assign the CP to the out-of-plane \Ag vibration. By performing a Gaussian fitting procedure for the positive PB signal corresponding to the A' and B' contributions, as shown in Figure~\ref{fig:dynamics}e, we can obtain the temporal dynamics of the peak energies (Figure~\ref{fig:dynamics}f). Both PB peaks are initially red-shifted by the photoexcitation and follow a similar recovery dynamics. Interestingly, the peak energies of both A' and B' resonances are also modulated (with matching phase) by the \Ag vibration with an amplitude of $\sim$2.5 meV or 5 meV peak-to-peak, as highlighted in the inset of Figure~\ref{fig:dynamics}f.

For the excitation mechanism of CPs, two main processes are customarily invoked; impulsive stimulated Raman scattering (ISRS)\cite{Dhar1994}, and displacive excitation of coherent phonons (DECP)\cite{Zeiger1992}. In ISRS, the CPs have a pump energy dependence that follows the excitation profile of the Raman tensor. The pump pulse transmits kinetic energy to the lattice atoms during a time interval much shorter than the oscillation period. At $t$ = 0, the atoms are in a quasi-equilibrium position, resulting in a sine oscillation. In DECP, on the other hand, the population of excited states changes the potential energy surface and thus the quasi-equilibrium position of the lattice. Therefore, at $t$ = 0 the lattice is at a maximum or minimum of the oscillating nuclear coordinate, resulting in a cosine oscillation. The high time resolution of our experiment allows a precise determination of the phase of the oscillatory \dR component and the identification of the cosine oscillation characteristic of DECP (see Supporting Information, Figure~S2). Moreover, we find that the oscillations in \MoTe have similar magnitudes, or slightly enhanced, for excitation at 2.36 eV compared to 1.91 eV (see Supporting Information, Figure~S4), while the Raman peak at 170 \cm is much weaker for excitation energy at 2.33 eV, corroborating the identification of different excitation mechanisms for Raman scattering and CPs.

\begin{figure}
    \includegraphics[width=1.0\linewidth]{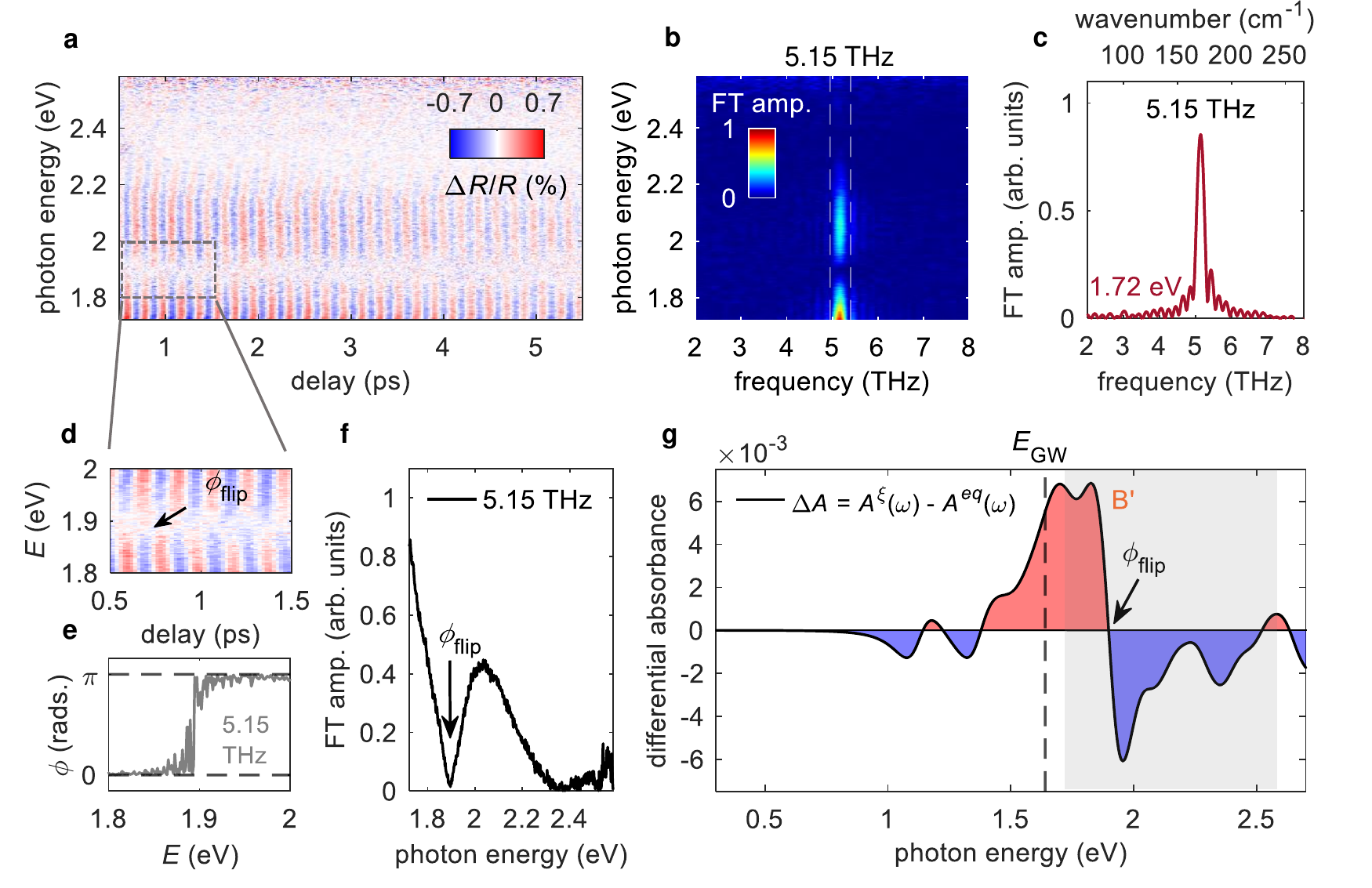}
    \caption{\textbf{Frequency analysis of coherent phonon oscillations in monolayer 2$H$-MoTe$_{2}$.} (\textbf{a}) Coherent component of the transient signal map, \dR. (\textbf{b}) Fourier transform (FT) map of the data in panel a. (\textbf{c)} FT frequency spectrum extracted at 1.72 eV, showing a single peak around 5.15 THz ($\sim$ 172 \cm). (\textbf{d}) Coherent component map showing the energy range (1.8 - 2.0 eV) near to the phase flip, labelled $\phi_{\rm flip}$. (\textbf{e}) FT phase spectrum confirming the phase flip from 0 to $\pi$ which occurs at $\sim$1.89 eV. (\textbf{f}) FT amplitude spectrum extracted at 5.15 THz (vertical dashed area in panel b). (\textbf{g}) Differential absorbance spectrum, $A^{\xi}(\omega) - A^{eq}(\omega)$ from \textit{ab initio} calculations. The vertical dashed line indicates the GW direct bandgap energy, while the orange label shows the position of the theoretical B' exciton transition. The grey shaded area highlights the experimentally explored energy range.}
  \label{fig:frequency}
\end{figure}

The spectral window where CP modulation of \dR is visible, as seen in Figure~\ref{fig:frequency}a,b, can be divided into two regions: above and below 1.89 eV, where the oscillations have opposite sign equivalent to a phase difference by $\pi$. The Fourier transform of the CP in Figure~\ref{fig:frequency}b,c, shows a single mode, which has constant frequency of 5.15 THz over the entire probe window. Its amplitude however, as shown in Figure~\ref{fig:frequency}f, changes dramatically over the range 1.7 to 2.3 eV, with a maximum at or below the lowest probe energy (1.72 eV), and a zero around 1.89 eV associated with a $\pi$ phase flip occurring at this energy, which is directly visible in the data in Figure~\ref{fig:frequency}d, and confirmed by Fourier analysis in Figure~\ref{fig:frequency}e.

To gain further insight into the probe energy dependence of the amplitude and phase of the 5.15 THz vibration, we calculated the change in the absorption spectrum caused by out-of-plane displacement of the Te atoms around the central Mo along the $c$-axis, mimicking an \Ag-type oscillation launched by the pulse. Starting from the result of the BSE calculations, as described in the Methods section, we obtain the polarizability per unit area, $\alpha_{2D}(\omega)$, and define an effective dielectric tensor, $\epsilon_{2D}(\omega)$ which takes into account the effects of quantum confinement in two-dimensions $\epsilon_{2D}(\omega) = 1 + \frac{4 \pi \alpha_{2D}(\omega)}{\Delta z}$ \cite{Molina-Sanchez2020}, where $\Delta z$ is the material thickness. In the case of monolayer MoTe$_{2}$, the theoretical value is $\Delta z = 7.66 $ \AA. The dielectric tensor is used to compute the absorbance according to $A(\omega) = \frac{\omega}{c} \epsilon_{2D}(\omega)$ \cite{Benardi2013}. The differential absorbance, $\Delta A$ was then calculated by subtracting the absorbance of the equilibrium structure, A$^{eq}$ from the absorbance, A$^{\xi}(\alpha)$ with atoms displaced by a fixed amount, $\alpha$ along the A$_{1g}$ phonon mode, $\xi$. The differential absorbance measured experimentally is of the order of 10$^{-2}$, and can be described in terms of a linear dependence on the atomic displacement $\alpha$, i.e. $A^{\xi}(\alpha) \simeq A^{eq} + \partial_{\alpha} A^{\xi} \alpha$. In the numerical simulations, $\alpha$ was fixed to obtain a stretching along the $c$-axis equivalent to 0.5\% of the Mo-Te bond length, measured from the center of the atoms. The value of $\partial_{\alpha}A^{\xi}$ was then extracted and used to estimate the much smaller atomic displacement (theoretical value $\alpha = 8.86 \times 10^{-3}$ \r{A}) corresponding to the signal measured experimentally.

The result, shown in Figure~\ref{fig:frequency}g, reproduces both the Fourier spectrum and the phase flip remarkably across the experimental energy window (grey shaded area). The calculated spectrum reveals a rich structure consisting of multiple optical transitions, where the peak widths have been inferred. Most notably, the spectrum is dominated by a large contribution centered around $\sim$ 2 eV which corresponds to the B' exciton absorption, suggesting a particularly strong coupling of this transition with the \Ag mode. The theoretical phase flip energy was found to be 1.90 eV, in almost perfect agreement with that found experimentally.

\begin{figure}
    \includegraphics[width=0.7\linewidth]{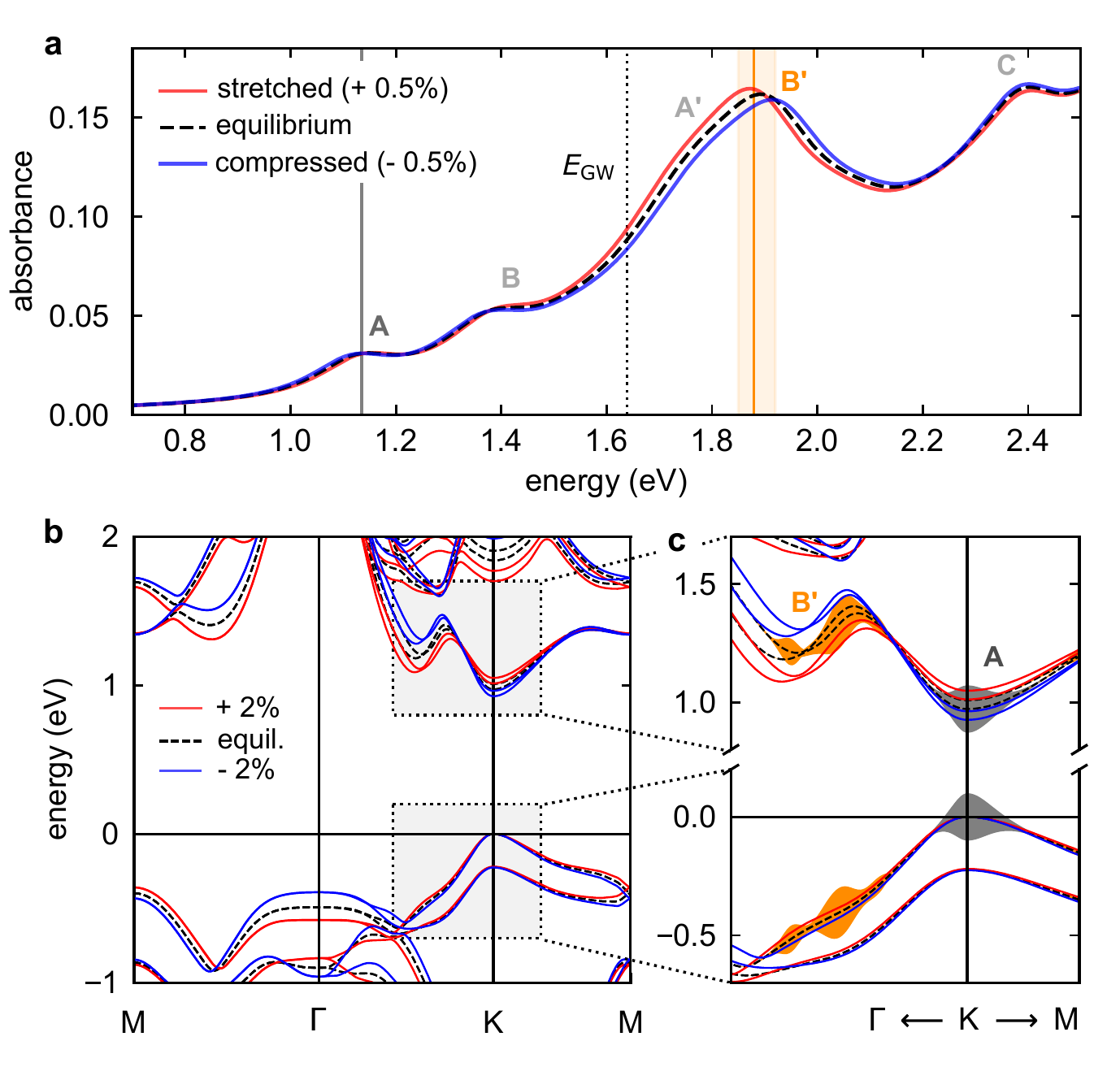}
    \caption{\textbf{Optical and electronic properties of monolayer 2$H$-MoTe$_{2}$ from \textit{ab initio} calculations.} (\textbf{a}) Optical absorption spectra for vertical stretching (red) and compression (blue) along the $c$-axis direction of the Mo-Te bond length equivalent to 0.5\%. The spectrum of the equilibrium structure (dashed line), i.e. no stretching or compression, is shown for comparison. The vertical lines indicate the A and B' exciton resonances and the GW direct bandgap energy, $E_{\mathrm{GW}}$. (\textbf{b}) Electronic band structure for 2\% stretching (red) and compression (blue). (\textbf{c}) Selected region around the $K$-point. The shaded areas represent the energy-momentum distribution of the optical transitions related to the A and B' excitons for the equilibrium structure.}
  \label{fig:theory}
\end{figure}

We now inspect the effects of \Ag mode atomic displacement on the electronic and optical properties of \MoTe in more detail. Figure~\ref{fig:theory}a shows calculations of the optical absorption of a monolayer in equilibrium, including excitonic effects, compared to a stretching (red line) and compression (blue line) of 0.5\% along the $c$-axis. The main features (labelled in Figure~\ref{fig:theory}a) are reproduced with excellent agreement to previous works\cite{Wilson1969,Ruppert2014} starting from the lowest energy A exciton at $\sim$ 1.1 eV. In particular, we emphasize the position of the B' transition around 2 eV, which is located within the continuum beyond the GW direct bandgap, and hence a large number of transitions (orange shaded area) have been taken into account to analyse it correctly. The optical absorption is subtly different except for the spectral region 1.6 to 2.1 eV, where there is dramatic rearrangement in both energy and magnitude. For out-of-plane compression, the A’ and B’ resonances shift to higher energies, with the B’ peak shifting more significantly, which leads to a larger separation of the two overlapping peaks, while upon stretching they shift to lower energies, increasing the overlap and forming a single, more intense peak. Consistently with our experimental results, the strongest modulation of the absorption occurs on the low energy side of the double peak, with a relative change of approximately 10\% for 0.5\% out-of-plane displacement.

The particularly strong coupling of the B' exciton with the \Ag mode leads to an energy shift of the absorption around the equilibrium position as the atoms oscillate. Such an energy modulation was observed in the experiments (Figure~\ref{fig:dynamics}f), as discussed previously. In the displacive excitation of coherent phonons, photoexcitation changes the nuclear quasi-equilibrium positions and at $t$ = 0 the nuclei are displaced relative to this new quasi-equilibrium. Their oscillation results in a periodic modulation of the absorption spectrum. Following the oscillatory component at energies below 1.89 eV, we observe a positive \dR at $t$ = 0. Therefore, at $t$ = 0 the spectrum is blue shifted relative to its new quasi-equilibrium, meaning the out-of-plane positions of the Te atoms are at a minimum. Their new quasi-equilibrium position is at a larger distance than prior to photoexcitation. Above 1.89 eV the same blue shift results in an increase of the absorption and hence an oscillatory modulation of opposite sign, i.e. a phase shift of $\pi$. 

To understand why the optical absorption exhibits such a dramatic change in a specific energy range, we now analyse the calculated electronic band structures, shown in Figure~\ref{fig:theory}b,c, for a larger displacement of 2\% in order to emphasize the effects. The calculations confirm the two valence band maxima and conduction band minima at the $K$ / $K$' point, which give rise to the lowest excitonic transitions A and B, outside the spectral window of our experiment. The A’ and B’ transitions, which appear as a broad double PB peak centred around 1.9 eV in the experiment, originate from momenta in the region around K, where the latter is found approximately halfway along the $K - \Gamma$ direction (see Figure~\ref{fig:theory}c). We find that the atomic displacement results in considerable modification of the bands between $K$ and $\Gamma$, especially close to the local minima of the conduction band. At the $K$-point, the dominant change in the electronic structure is bandgap renormalization which results in a comparatively small energy shift of the A exciton transition. However, there is particularly significant modification halfway along the $K - \Gamma$ where contributions of the B$'$ exciton are most important. The result is a large energy shift and change in the magnitude of the optical absorption in the spectral range close to this transition, as seen in Figure~\ref{fig:theory}a. Hence this explains the extremely intense coherent oscillations observed in experiment as the result of a modulation of the optical absorption due to the out-of-plane atomic displacement launched by the optical pulse.


\section{Conclusion}

We investigated the generation and detection of coherent phonons in monolayer 2$H$-\MoTe using a combination of femtosecond pump-probe microscopy and \textit{ab initio} calculations. In excellent agreement between experiment and theory, we found that photoexcitation stimulates the out-of-plane \Ag vibration, which strongly modulates the absorption in the visible range, especially around the A’ and B’ excitons. We identified a displacive excitation mechanism, where photoexcitation shifts the quasi-equilibrium positions of the Te atoms to a larger out-of-plane distance. Our calculations also predict modulations of significant magnitude around the A and B excitons, thus expanding the potential for the coherent control of optical phonons and excitons, via optical excitation or applied out-of-plane compressive strain, into the near-infrared region down to $\sim$1 eV.


\section{Methods}

\subsection{Experimental}

Ultrafast spectroscopy experiments were performed using a custom broadband transient absorption microscope setup based on a Ti:sapphire laser system (Coherent Libra) which outputs 100 fs pulses at 1.55 eV with 2 kHz repetition rate. The pump was generated using a non-collinear optical parametric amplifier (NOPA) tuned to a centre energy of $\sim$ 2.36 eV. The probe was created by white light continuum generation in a 1 mm sapphire plate using the laser fundamental. Pump and probe beams were focused onto the sample with a diameter of $\sim$ 5 and 3 \um, respectively, using an achromatic objective lens. Cross polarization was used to avoid scattering artefacts. The pump was pre-compressed using chirp mirrors to account for all transmissive optical elements in the beam path that introduce dispersion. The pulse compression was optimized by measuring a target metallic sample with a nearly instantaneous response, resulting in an overall temporal resolution of $\leq$ 40 fs, comparable to the build-up time of excitons in MoS$_2$ as measured previously\cite{Trovatello2020b}. A pump fluence of $\sim$ 500 \uJcm was used. The sample was mounted inside an ultra low vibration closed-cycle cryostat, and the temperature maintained at $T$ = 10 K. The setup operates in backscattering geometry, whereby the reflected probe is dispersed by a spectrometer onto a CCD camera, and the differential reflectance of the sample (\dR) is measured as a function of pump-probe delay.

\subsection{Computational}

Theoretical investigation of the absorbance has been carried out by means of \textit{ab initio} simulations. DFT\cite{HohenbergKohn1964,KohnSham1965} and DFPT \cite{Baroni2001} calculations were performed with \textit{QUANTUM ESPRESSO} \cite{Giannozzi2009,Giannozzi2017,Giannozzi2020}, where the Perdew–Burke–Ernzerhof exchange-correlation  functional \cite{Perdew1996} and a plane-wave cutoff at 50 Ha were used. To replace the effect of core electrons, fully relativistic pseudopotentials were used from PSEUDODOJO \cite{vanSetten2018} since spin-orbit coupling is taken into account. A vacuum distance of 40 a.u. was imposed in order to avoid the fictitious interactions between periodic layers. To calculate the quasiparticle correction, the GW method \cite{Onida2002,Reining2018} was used in single-shot mode G$_{0}$W$_{0}$, applying the plasmon-pole approximation. In order to compute the optical properties taking into account the excitonic effects, the Bethe-Salpeter equation (BSE) \cite{Marsili2021} was solved, where the Tamm-Dancoff approximation was applied. 14 valence bands and 13 conduction bands were considered for the correct description of the optical absorption. The polarizability per unit area of the \MoTe monolayer was obtained from the BSE results, and used to calculate the effective dielectric function and the absorbance. In order to compute the coherent part of the differential absorbance, DFT and BSE simulations were first performed for equilibrium atomic positions and then with the atoms displaced along the Raman active \Ag phonon eigen-mode $\xi$, as explained in Ref \cite{Miranda2017}. Both GW and BSE simulations were performed using \textit{YAMBO} code \cite{Sangalli2019}. All calculations were performed with a $42 \times 42 \times 1$ Monkhorst-Pack grid \cite{Monkhorst1976} for the Brillouin zone.

\begin{acknowledgement}

This work has received funding from the European Union's Horizon 2020 research and innovation programme under grant agreement 881603. A.G. acknowledges support by the European Union Marie Sklodowska-Curie Actions (project ENOSIS H2020-MSCA-IF-2020-101029644). C.T. acknowledges the Optica Foundation and Coherent Inc. for supporting her research through the Bernard J. Couillaud Prize. A.M.-S. acknowledges the Ram\'on y Cajal programme (grant RYC2018-024024-I; MINECO, Spain), Generalitat Valenciana, program SEJIGENT (reference 2021/034), project Magnons in magnetic 2D materials for a novel electronics (2D MAGNONICS), and Planes complementarios de I+D+I en materiales avanzados, project SPINO2D, reference MFA/2022/009. A.M.-S. and J.C.-V. acknowledges the Spanish Ministry of Science MICINN (project PID2020-112507GB-I00, Novel quantum states in heterostructures of 2D materials and Grant PRE2021-097581). D.S. acknowledges PRIN BIOX Grant No. 20173B72NB from MIUR (Italy), European Union project MaX Materials design at the eXascale H2020-EINFRA-2015-1 (Grants Agreement No. 824143), and COST Action TUMIEE CA17126, supported by COST (European Cooperation in Science and Technology).

\end{acknowledgement}

\newpage

\section{Figures}

\newpage

\bibliography{refs}

\providecommand{\latin}[1]{#1}
\makeatletter
\providecommand{\doi}
  {\begingroup\let\do\@makeother\dospecials
  \catcode`\{=1 \catcode`\}=2 \doi@aux}
\providecommand{\doi@aux}[1]{\endgroup\texttt{#1}}
\makeatother
\providecommand*\mcitethebibliography{\thebibliography}
\csname @ifundefined\endcsname{endmcitethebibliography}
  {\let\endmcitethebibliography\endthebibliography}{}
\begin{mcitethebibliography}{76}
\providecommand*\natexlab[1]{#1}
\providecommand*\mciteSetBstSublistMode[1]{}
\providecommand*\mciteSetBstMaxWidthForm[2]{}
\providecommand*\mciteBstWouldAddEndPuncttrue
  {\def\EndOfBibitem{\unskip.}}
\providecommand*\mciteBstWouldAddEndPunctfalse
  {\let\EndOfBibitem\relax}
\providecommand*\mciteSetBstMidEndSepPunct[3]{}
\providecommand*\mciteSetBstSublistLabelBeginEnd[3]{}
\providecommand*\EndOfBibitem{}
\mciteSetBstSublistMode{f}
\mciteSetBstMaxWidthForm{subitem}{(\alph{mcitesubitemcount})}
\mciteSetBstSublistLabelBeginEnd
  {\mcitemaxwidthsubitemform\space}
  {\relax}
  {\relax}

\bibitem[Bardeen and Shockley(1950)Bardeen, and Shockley]{Bardeen1950}
Bardeen,~J.; Shockley,~W. Deformation Potentials and Mobilities in Non-Polar
  Crystals. \emph{Phys. Rev.} \textbf{1950}, \emph{80}, 72, DOI:
  \doi{10.1103/PhysRev.80.72}\relax
\mciteBstWouldAddEndPuncttrue
\mciteSetBstMidEndSepPunct{\mcitedefaultmidpunct}
{\mcitedefaultendpunct}{\mcitedefaultseppunct}\relax
\EndOfBibitem
\bibitem[Fischetti and Laux(1996)Fischetti, and Laux]{Fischetti1996}
Fischetti,~M.~V.; Laux,~S.~E. Band Structure, Deformation Potentials, and
  Carrier Mobility in Strained Si, Ge, and SiGe Alloys. \emph{J. Appl. Phys.}
  \textbf{1996}, \emph{80}, 2234, DOI: \doi{10.1063/1.363052}\relax
\mciteBstWouldAddEndPuncttrue
\mciteSetBstMidEndSepPunct{\mcitedefaultmidpunct}
{\mcitedefaultendpunct}{\mcitedefaultseppunct}\relax
\EndOfBibitem
\bibitem[Gunst \latin{et~al.}(2016)Gunst, Markussen, Stokbro, and
  Brandbyge]{Gunst2016}
Gunst,~T.; Markussen,~T.; Stokbro,~K.; Brandbyge,~M. First-principles method
  for electron-phonon coupling and electron mobility: Applications to
  two-dimensional materials. \emph{Phys. Rev. B} \textbf{2016}, \emph{93},
  035414, DOI: \doi{10.1103/PhysRevB.93.035414}\relax
\mciteBstWouldAddEndPuncttrue
\mciteSetBstMidEndSepPunct{\mcitedefaultmidpunct}
{\mcitedefaultendpunct}{\mcitedefaultseppunct}\relax
\EndOfBibitem
\bibitem[Allen(1987)]{Allen1987}
Allen,~P.~B. Theory of Thermal Relaxation of Electrons in Metals. \emph{Phys.
  Rev. Lett.} \textbf{1987}, \emph{59}, 1460, DOI:
  \doi{10.1103/PhysRevLett.59.1460}\relax
\mciteBstWouldAddEndPuncttrue
\mciteSetBstMidEndSepPunct{\mcitedefaultmidpunct}
{\mcitedefaultendpunct}{\mcitedefaultseppunct}\relax
\EndOfBibitem
\bibitem[Gadermaier \latin{et~al.}(2010)Gadermaier, Alexandrov, Kabanov, Kusar,
  Mertelj, Yao, Manzoni, Brida, Cerullo, and Mihailovic]{Gadermaier2010}
Gadermaier,~C.; Alexandrov,~A.~S.; Kabanov,~V.~V.; Kusar,~P.; Mertelj,~T.;
  Yao,~X.; Manzoni,~C.; Brida,~D.; Cerullo,~G.; Mihailovic,~D. Electron-Phonon
  Coupling in High-Temperature Cuprate Superconductors Determined from Electron
  Relaxation Rates. \emph{Phys. Rev. Lett.} \textbf{2010}, \emph{105}, 1, DOI:
  \doi{10.1103/PhysRevLett.105.257001}\relax
\mciteBstWouldAddEndPuncttrue
\mciteSetBstMidEndSepPunct{\mcitedefaultmidpunct}
{\mcitedefaultendpunct}{\mcitedefaultseppunct}\relax
\EndOfBibitem
\bibitem[He \latin{et~al.}(2020)He, Chebl, and Yang]{He2020}
He,~X.; Chebl,~M.; Yang,~D.-S. Cross-Examination of Ultrafast Structural,
  Interfacial, and Carrier Dynamics of Supported Monolayer MoS$_2$. \emph{Nano
  Lett.} \textbf{2020}, \emph{20}, 2026--2033, DOI:
  \doi{10.1021/acs.nanolett.9b05344}\relax
\mciteBstWouldAddEndPuncttrue
\mciteSetBstMidEndSepPunct{\mcitedefaultmidpunct}
{\mcitedefaultendpunct}{\mcitedefaultseppunct}\relax
\EndOfBibitem
\bibitem[Molina-S{\'a}nchez \latin{et~al.}(2017)Molina-S{\'a}nchez, Sangalli,
  Wirtz, and Marini]{Molina-Sanchez2017}
Molina-S{\'a}nchez,~A.; Sangalli,~D.; Wirtz,~L.; Marini,~A. Ab initio
  calculations of ultrashort carrier dynamics in twodimensional materials:
  Valley depolarization in single-layer WSe$_2$. \emph{Nano Lett.}
  \textbf{2017}, \emph{17}, 4549, DOI: \doi{10.1021/acs.nanolett.7b00175}\relax
\mciteBstWouldAddEndPuncttrue
\mciteSetBstMidEndSepPunct{\mcitedefaultmidpunct}
{\mcitedefaultendpunct}{\mcitedefaultseppunct}\relax
\EndOfBibitem
\bibitem[Selig \latin{et~al.}(2016)Selig, Berghuser, Raja, Nagler, Schller,
  Heinz, Korn, Chernikov, Malic, and Knorr]{Selig2016}
Selig,~M.; Berghuser,~G.; Raja,~A.; Nagler,~P.; Schller,~C.; Heinz,~T.;
  Korn,~T.; Chernikov,~A.; Malic,~E.; Knorr,~A. Excitonic Linewidth and
  Coherence Lifetime in Monolayer Transition Metal Dichalcogenides. \emph{Nat.
  Comms.} \textbf{2016}, \emph{7}, 13279, DOI: \doi{10.1038/ncomms13279}\relax
\mciteBstWouldAddEndPuncttrue
\mciteSetBstMidEndSepPunct{\mcitedefaultmidpunct}
{\mcitedefaultendpunct}{\mcitedefaultseppunct}\relax
\EndOfBibitem
\bibitem[Porer \latin{et~al.}(2014)Porer, Leierseder, M{\'e}nard, Dachraoui,
  Mouchliadis, Perakis, Heinzmann, Demsar, Rossnagel, and Huber]{Porer2014}
Porer,~M.; Leierseder,~U.; M{\'e}nard,~J.-M.; Dachraoui,~H.; Mouchliadis,~L.;
  Perakis,~I.~E.; Heinzmann,~U.; Demsar,~J.; Rossnagel,~K.; Huber,~R.
  Non-thermal separation of electronic and structural orders in a persisting
  charge density wave. \emph{Nat. Mater.} \textbf{2014}, \emph{13}, 857, DOI:
  \doi{10.1038/nmat4042}\relax
\mciteBstWouldAddEndPuncttrue
\mciteSetBstMidEndSepPunct{\mcitedefaultmidpunct}
{\mcitedefaultendpunct}{\mcitedefaultseppunct}\relax
\EndOfBibitem
\bibitem[Hedayat \latin{et~al.}(2019)Hedayat, Sayers, Bugini, Dallera,
  Wolverson, Batten, Karbassi, Friedemann, Cerullo, van Wezel, Clark, Carpene,
  and Da~Como]{Hedayat2019}
Hedayat,~H.; Sayers,~C.~J.; Bugini,~D.; Dallera,~C.; Wolverson,~D.; Batten,~T.;
  Karbassi,~S.; Friedemann,~S.; Cerullo,~G.; van Wezel,~J.; Clark,~S.~R.;
  Carpene,~E.; Da~Como,~E. Excitonic and lattice contributions to the charge
  density wave in $1T$-TiSe$_2$ revealed by a phonon bottleneck. \emph{Phys.
  Rev. Res.} \textbf{2019}, \emph{1}, 023029, DOI:
  \doi{10.1103/PhysRevResearch.1.023029}\relax
\mciteBstWouldAddEndPuncttrue
\mciteSetBstMidEndSepPunct{\mcitedefaultmidpunct}
{\mcitedefaultendpunct}{\mcitedefaultseppunct}\relax
\EndOfBibitem
\bibitem[Maklar \latin{et~al.}(2021)Maklar, Windsor, Nicholson, Puppin,
  Walmsley, Esposito, Porer, Rittmann, Leuenberger, Kubli, Savoini, Abreu,
  Johnson, Beaud, Ingold, Staub, Fisher, Ernstorfer, Wolf, and
  Rettig]{Maklar2021}
Maklar,~J. \latin{et~al.}  Nonequilibrium charge-density-wave order beyond the
  thermal limit. \emph{Nat. Comms.} \textbf{2021}, \emph{12}, 2499, DOI:
  \doi{10.1038/s41467-021-22778-w}\relax
\mciteBstWouldAddEndPuncttrue
\mciteSetBstMidEndSepPunct{\mcitedefaultmidpunct}
{\mcitedefaultendpunct}{\mcitedefaultseppunct}\relax
\EndOfBibitem
\bibitem[Sayers \latin{et~al.}(2022)Sayers, Dal~Conte, Wolverson, Gadermaier,
  Cerullo, Carpene, and Da~Como]{Sayers2022}
Sayers,~C.~J.; Dal~Conte,~S.; Wolverson,~D.; Gadermaier,~C.; Cerullo,~G.;
  Carpene,~E.; Da~Como,~E. Spectrally Resolving the Phase and Amplitude of
  Coherent Phonons in the Charge Density Wave State of 1$T$-TaSe$_2$.
  \emph{Adv. Optical Mater} \textbf{2022}, \emph{10}, 62, DOI:
  \doi{10.1002/adom.202200362}\relax
\mciteBstWouldAddEndPuncttrue
\mciteSetBstMidEndSepPunct{\mcitedefaultmidpunct}
{\mcitedefaultendpunct}{\mcitedefaultseppunct}\relax
\EndOfBibitem
\bibitem[Cerullo and Manzoni(2007)Cerullo, and Manzoni]{Cerullo2007}
Cerullo,~G.; Manzoni,~C. \emph{Coherent Vibrational Dynamics}; CRC Press: Boca
  Raton, 2007; Chapter Time-Domain Vibrational Spectroscopy: Principle and
  Experimental Tools, pp 1--48\relax
\mciteBstWouldAddEndPuncttrue
\mciteSetBstMidEndSepPunct{\mcitedefaultmidpunct}
{\mcitedefaultendpunct}{\mcitedefaultseppunct}\relax
\EndOfBibitem
\bibitem[Dekorsy \latin{et~al.}(2010)Dekorsy, Cho, and Kurz]{Dekorsy2010}
Dekorsy,~T.; Cho,~G.~C.; Kurz,~H. \emph{Light Scattering in Solids VIII};
  Springer Berlin Heidelberg, 2010; pp 169--209\relax
\mciteBstWouldAddEndPuncttrue
\mciteSetBstMidEndSepPunct{\mcitedefaultmidpunct}
{\mcitedefaultendpunct}{\mcitedefaultseppunct}\relax
\EndOfBibitem
\bibitem[Ishioka and Misochko(2010)Ishioka, and Misochko]{Ishioka2010}
Ishioka,~K.; Misochko,~O.~V. Progress in Ultrafast Intense Laser Science: Vol
  V. 2010; pp 47--63\relax
\mciteBstWouldAddEndPuncttrue
\mciteSetBstMidEndSepPunct{\mcitedefaultmidpunct}
{\mcitedefaultendpunct}{\mcitedefaultseppunct}\relax
\EndOfBibitem
\bibitem[Wang \latin{et~al.}(2011)Wang, Schoenlein, Peteanu, Mathies, and
  Shank]{Wang2011}
Wang,~Q.; Schoenlein,~R.~W.; Peteanu,~L.~A.; Mathies,~R.~A.; Shank,~C.~V.
  Vibrationally Coherent Photochemistry in the Femtosecond Primary Event of
  Vision. \emph{Science} \textbf{2011}, \emph{266}, 422--424, DOI:
  \doi{10.1126/science.7939680}\relax
\mciteBstWouldAddEndPuncttrue
\mciteSetBstMidEndSepPunct{\mcitedefaultmidpunct}
{\mcitedefaultendpunct}{\mcitedefaultseppunct}\relax
\EndOfBibitem
\bibitem[Feldmann \latin{et~al.}(1992)Feldmann, Leo, Shah, Miller, Cunningham,
  Meier, Plessen, Schulze, Thomas, Schmitt-Rink, and Optical]{Feldmann1992}
Feldmann,~J.; Leo,~K.; Shah,~J.; Miller,~D. A.~B.; Cunningham,~J.~E.;
  Meier,~T.; Plessen,~V.; Schulze,~G.; Thomas,~A.; Schmitt-Rink,~P.;
  Optical,~S. Investigation of Bloch oscillations in a semiconductor
  superlattice. \emph{Phys. Rev. B} \textbf{1992}, \emph{46}, 7252--7255, DOI:
  \doi{10.1007/978-3-642-84910-7_146}\relax
\mciteBstWouldAddEndPuncttrue
\mciteSetBstMidEndSepPunct{\mcitedefaultmidpunct}
{\mcitedefaultendpunct}{\mcitedefaultseppunct}\relax
\EndOfBibitem
\bibitem[Dekorsy \latin{et~al.}(1995)Dekorsy, Auer, Waschke, Bakker, Roskos,
  Kurz, Wagner, and Grosse]{Dekorsy1995}
Dekorsy,~T.; Auer,~H.; Waschke,~C.; Bakker,~H.~J.; Roskos,~H.~G.; Kurz,~H.;
  Wagner,~V.; Grosse,~P. Emission of Submillimeter Electromagnetic Waves by
  Coherent Phonons. \emph{Phys. Rev. Lett.} \textbf{1995}, \emph{74}, 738--741,
  DOI: \doi{10.1103/PhysRevLett.74.738}\relax
\mciteBstWouldAddEndPuncttrue
\mciteSetBstMidEndSepPunct{\mcitedefaultmidpunct}
{\mcitedefaultendpunct}{\mcitedefaultseppunct}\relax
\EndOfBibitem
\bibitem[Li and Zhu(2013)Li, and Zhu]{Li2013}
Li,~J.~J.; Zhu,~K.~D. All-Optical Mass Sensing with Coupled Mechanical
  Resonator Systems. \emph{Phys. Rep.} \textbf{2013}, \emph{525}, 223, DOI:
  \doi{10.1016/j.physrep.2012.11.003}\relax
\mciteBstWouldAddEndPuncttrue
\mciteSetBstMidEndSepPunct{\mcitedefaultmidpunct}
{\mcitedefaultendpunct}{\mcitedefaultseppunct}\relax
\EndOfBibitem
\bibitem[Lanzillotti-Kimura \latin{et~al.}(2018)Lanzillotti-Kimura, O’Brien,
  Rho, Suchowski, Yin, and Zhang]{Lanzillotti-Kimura2018}
Lanzillotti-Kimura,~N.~D.; O’Brien,~K.~P.; Rho,~J.; Suchowski,~H.; Yin,~X.;
  Zhang,~X. Polarization-Controlled Coherent Phonon Generation in
  Acoustoplasmonic Metasurfaces. \emph{Phys. Rev. B} \textbf{2018}, \emph{97},
  23540, DOI: \doi{10.1103/PhysRevB.97.235403}\relax
\mciteBstWouldAddEndPuncttrue
\mciteSetBstMidEndSepPunct{\mcitedefaultmidpunct}
{\mcitedefaultendpunct}{\mcitedefaultseppunct}\relax
\EndOfBibitem
\bibitem[Baldini \latin{et~al.}(2018)Baldini, Palmieri, Dominguez, Ruello,
  Rubio, and Chergui]{Baldini2018}
Baldini,~E.; Palmieri,~T.; Dominguez,~A.; Ruello,~P.; Rubio,~A.; Chergui,~M.
  Phonon-Driven Selective Modulation of Exciton Oscillator Strengths in Anatase
  TiO2 Nanoparticles. \emph{Nano Lett.} \textbf{2018}, \emph{18}, 5007--5014,
  DOI: \doi{10.1021/acs.nanolett.8b01837}\relax
\mciteBstWouldAddEndPuncttrue
\mciteSetBstMidEndSepPunct{\mcitedefaultmidpunct}
{\mcitedefaultendpunct}{\mcitedefaultseppunct}\relax
\EndOfBibitem
\bibitem[Baldini \latin{et~al.}(2019)Baldini, Dominguez, Palmieri, Cannelli,
  Rubio, Ruello, and Chergui]{Baldini2019}
Baldini,~E.; Dominguez,~A.; Palmieri,~T.; Cannelli,~O.; Rubio,~A.; Ruello,~P.;
  Chergui,~M. Exciton Control in a Room Temperature Bulk Semiconductor with
  Coherent Strain Pulses. \emph{Sci. Adv.} \textbf{2019}, \emph{5}, DOI:
  \doi{10.1126/sciadv.aax2937}\relax
\mciteBstWouldAddEndPuncttrue
\mciteSetBstMidEndSepPunct{\mcitedefaultmidpunct}
{\mcitedefaultendpunct}{\mcitedefaultseppunct}\relax
\EndOfBibitem
\bibitem[Hase \latin{et~al.}(2013)Hase, Katsuragawa, Constantinescu, and
  Petek]{Hase2013}
Hase,~M.; Katsuragawa,~M.; Constantinescu,~A.~M.; Petek,~H. Coherent
  Phonon-Induced Optical Modulation in Semiconductors at Terahertz Frequencies.
  \emph{New J. Phys.} \textbf{2013}, \emph{15}, 055018, DOI:
  \doi{10.1088/1367-2630/15/5/055018}\relax
\mciteBstWouldAddEndPuncttrue
\mciteSetBstMidEndSepPunct{\mcitedefaultmidpunct}
{\mcitedefaultendpunct}{\mcitedefaultseppunct}\relax
\EndOfBibitem
\bibitem[Luer \latin{et~al.}(2009)Luer, Gadermaier, Crochet, Hertel, Brida, and
  Lanzani]{Luer2009}
Luer,~L.; Gadermaier,~C.; Crochet,~J.; Hertel,~T.; Brida,~D.; Lanzani,~G.
  Coherent Phonon Dynamics in Semiconducting Carbon Nanotubes: A Quantitative
  Study of Electron-Phonon Coupling. \emph{Phys. Rev. Lett.} \textbf{2009},
  \emph{102}, 1, DOI: \doi{10.1103/PhysRevLett.102.127401}\relax
\mciteBstWouldAddEndPuncttrue
\mciteSetBstMidEndSepPunct{\mcitedefaultmidpunct}
{\mcitedefaultendpunct}{\mcitedefaultseppunct}\relax
\EndOfBibitem
\bibitem[Ni \latin{et~al.}(2017)Ni, Huynh, Cheminal, Thomas, Shivanna,
  Hinrichsen, Ahmad, Sadhanala, and Rao]{Ni2017}
Ni,~L.~M.; Huynh,~U.; Cheminal,~A.; Thomas,~T.~H.; Shivanna,~R.;
  Hinrichsen,~T.~F.; Ahmad,~S.; Sadhanala,~A.; Rao,~A. Real-Time Observation of
  Exciton-Phonon Coupling Dynamics in Self-Assembled Hybrid Perovskite Quantum
  Wells. \emph{ACS Nano} \textbf{2017}, \emph{11}, 10834--10843, DOI:
  \doi{10.1021/acsnano.7b03984}\relax
\mciteBstWouldAddEndPuncttrue
\mciteSetBstMidEndSepPunct{\mcitedefaultmidpunct}
{\mcitedefaultendpunct}{\mcitedefaultseppunct}\relax
\EndOfBibitem
\bibitem[Trovatello \latin{et~al.}(2020)Trovatello, Miranda, Molina-Sánchez,
  Borrego-Varillas, Manzoni, Moretti, Ganzer, Maiuri, Wang, Dumcenco, Kis,
  Wirtz, Marini, Soavi, Ferrari, Cerullo, Sangalli, and
  Dal~Conte]{Trovatello2020a}
Trovatello,~C. \latin{et~al.}  Strongly coupled coherent phonons in
  single-layer MoS$_2$. \emph{ACS Nano} \textbf{2020}, \emph{14}, 5700, DOI:
  \doi{10.1021/acsnano.0c00309}\relax
\mciteBstWouldAddEndPuncttrue
\mciteSetBstMidEndSepPunct{\mcitedefaultmidpunct}
{\mcitedefaultendpunct}{\mcitedefaultseppunct}\relax
\EndOfBibitem
\bibitem[Mor \latin{et~al.}(2021)Mor, Gosetti, Molina-Sanchez, Sangalli,
  Achilli, Agekyan, Franceschini, Giannetti, Sangaletti, and Pagliara]{Mor2021}
Mor,~S.; Gosetti,~V.; Molina-Sanchez,~A.; Sangalli,~D.; Achilli,~S.;
  Agekyan,~V.~F.; Franceschini,~P.; Giannetti,~C.; Sangaletti,~L.; Pagliara,~S.
  Photoinduced modulation of the excitonic resonance via coupling with coherent
  phonons in a layered semiconductor. \emph{Phys. Rev. Research} \textbf{2021},
  \emph{3}, 043175, DOI: \doi{10.1103/PhysRevResearch.3.043175}\relax
\mciteBstWouldAddEndPuncttrue
\mciteSetBstMidEndSepPunct{\mcitedefaultmidpunct}
{\mcitedefaultendpunct}{\mcitedefaultseppunct}\relax
\EndOfBibitem
\bibitem[Radisavljevic \latin{et~al.}(2011)Radisavljevic, Radenovic, Brivio,
  Giacometti, and Kis]{Radisavljevic2011}
Radisavljevic,~B.; Radenovic,~A.; Brivio,~J.; Giacometti,~V.; Kis,~A.
  Single-Layer MoS$_2$ Transistors. \emph{Nat. Nanotechnology} \textbf{2011},
  \emph{6}, 147, DOI: \doi{10.1038/nnano.2010.279}\relax
\mciteBstWouldAddEndPuncttrue
\mciteSetBstMidEndSepPunct{\mcitedefaultmidpunct}
{\mcitedefaultendpunct}{\mcitedefaultseppunct}\relax
\EndOfBibitem
\bibitem[Wang \latin{et~al.}(2012)Wang, Kalantar-Zadeh, Kis, Coleman, and
  Strano]{Wang2012}
Wang,~Q.~H.; Kalantar-Zadeh,~K.; Kis,~A.; Coleman,~J.~N.; Strano,~M.~S.
  Electronics and Optoelectronics of Two-Dimensional Transition Metal
  Dichalcogenides. \emph{Nat. Nanotechnology} \textbf{2012}, \emph{7}, 699,
  DOI: \doi{10.1038/nnano.2012.193}\relax
\mciteBstWouldAddEndPuncttrue
\mciteSetBstMidEndSepPunct{\mcitedefaultmidpunct}
{\mcitedefaultendpunct}{\mcitedefaultseppunct}\relax
\EndOfBibitem
\bibitem[Fiori \latin{et~al.}(2014)Fiori, Bonaccorso, Iannaccone, Palacios,
  Neumaier, Seabaugh, Banerjee, and Colombo]{Fiori2014}
Fiori,~G.; Bonaccorso,~F.; Iannaccone,~G.; Palacios,~T.; Neumaier,~D.;
  Seabaugh,~A.; Banerjee,~S.~K.; Colombo,~L. Electronics Based on
  Two-Dimensional Materials. \emph{Nat. Nanotechnology} \textbf{2014},
  \emph{9}, 768, DOI: \doi{10.1038/nnano.2014.207}\relax
\mciteBstWouldAddEndPuncttrue
\mciteSetBstMidEndSepPunct{\mcitedefaultmidpunct}
{\mcitedefaultendpunct}{\mcitedefaultseppunct}\relax
\EndOfBibitem
\bibitem[Jariwala \latin{et~al.}(2014)Jariwala, Sangwan, Lauhon, Marks, and
  Hersam]{Jariwala2014}
Jariwala,~D.; Sangwan,~V.~K.; Lauhon,~L.~J.; Marks,~T.~J.; Hersam,~M.~C.
  Emerging Device Applications for Semiconducting TwoDimensional Transition
  Metal Dichalcogenides. \emph{ACS Nano} \textbf{2014}, \emph{8}, 1102, DOI:
  \doi{10.1021/nn500064s}\relax
\mciteBstWouldAddEndPuncttrue
\mciteSetBstMidEndSepPunct{\mcitedefaultmidpunct}
{\mcitedefaultendpunct}{\mcitedefaultseppunct}\relax
\EndOfBibitem
\bibitem[Mak and Shan(2016)Mak, and Shan]{Mak2016}
Mak,~K.~F.; Shan,~J. Photonics and Optoelectronics of 2D Semiconductor
  Transition Metal Dichalcogenides. \emph{Nat. Photonics} \textbf{2016},
  \emph{10}, 216, DOI: \doi{10.1038/nphoton.2015.282}\relax
\mciteBstWouldAddEndPuncttrue
\mciteSetBstMidEndSepPunct{\mcitedefaultmidpunct}
{\mcitedefaultendpunct}{\mcitedefaultseppunct}\relax
\EndOfBibitem
\bibitem[Koppens \latin{et~al.}(2014)Koppens, Mueller, Avouris, Ferrari,
  Vitiello, and Polini]{Koppens2014}
Koppens,~F. L.~H.; Mueller,~T.; Avouris,~P.; Ferrari,~A.~C.; Vitiello,~M.~S.;
  Polini,~M. Photodetectors based on graphene, other two-dimensional materials
  and hybrid systems. \emph{Nat. Nanotechnology} \textbf{2014}, \emph{9}, 780,
  DOI: \doi{10.1038/nnano.2014.215}\relax
\mciteBstWouldAddEndPuncttrue
\mciteSetBstMidEndSepPunct{\mcitedefaultmidpunct}
{\mcitedefaultendpunct}{\mcitedefaultseppunct}\relax
\EndOfBibitem
\bibitem[Jeong \latin{et~al.}(2016)Jeong, Jin, Rhim, Debbichi, Park, Jang, Lee,
  Chae, Lee, Kim, Jung, and Yee]{Jeong2016}
Jeong,~T.~Y.; Jin,~B.~M.; Rhim,~S.~H.; Debbichi,~L.; Park,~J.; Jang,~Y.~D.;
  Lee,~H.~R.; Chae,~D.-H.; Lee,~D.; Kim,~Y.-H.; Jung,~S.; Yee,~K.~J. Coherent
  Lattice Vibrations in Mono- and Few-Layer WSe$_2$. \emph{ACS Nano}
  \textbf{2016}, \emph{10}, 5560, DOI: \doi{10.1021/acsnano.6b02253}\relax
\mciteBstWouldAddEndPuncttrue
\mciteSetBstMidEndSepPunct{\mcitedefaultmidpunct}
{\mcitedefaultendpunct}{\mcitedefaultseppunct}\relax
\EndOfBibitem
\bibitem[Cho \latin{et~al.}(2015)Cho, Kim, Kim, Zhao, Seok, Keum, Baik, Choe,
  Chang, Suenaga, Kim, Lee, and Yang]{Cho2015}
Cho,~S.; Kim,~S.; Kim,~J.~H.; Zhao,~J.; Seok,~J.; Keum,~D.~H.; Baik,~J.;
  Choe,~D.~H.; Chang,~K.~J.; Suenaga,~K.; Kim,~S.~W.; Lee,~Y.~H.; Yang,~H.
  Phase Patterning for Ohmic Homojunction Contact in MoTe$_2$. \emph{Science}
  \textbf{2015}, \emph{349}, 6248, DOI: \doi{10.1126/science.aab3175}\relax
\mciteBstWouldAddEndPuncttrue
\mciteSetBstMidEndSepPunct{\mcitedefaultmidpunct}
{\mcitedefaultendpunct}{\mcitedefaultseppunct}\relax
\EndOfBibitem
\bibitem[Keum \latin{et~al.}(2015)Keum, Cho, Kim, Choe, Sung, Kan, Kang, Hwang,
  Kim, Yang, Chang, and Lee]{Keum2015}
Keum,~D.~H.; Cho,~S.; Kim,~J.~H.; Choe,~D.~H.; Sung,~H.~J.; Kan,~M.; Kang,~H.;
  Hwang,~J.~Y.; Kim,~S.~W.; Yang,~H.; Chang,~K.~J.; Lee,~Y.~H. Bandgap Opening
  in Few-Layered Monoclinic MoTe$_2$. \emph{Nat. Phys.} \textbf{2015},
  \emph{11}, 6, DOI: \doi{10.1038/nphys3314}\relax
\mciteBstWouldAddEndPuncttrue
\mciteSetBstMidEndSepPunct{\mcitedefaultmidpunct}
{\mcitedefaultendpunct}{\mcitedefaultseppunct}\relax
\EndOfBibitem
\bibitem[Lee \latin{et~al.}(2018)Lee, Jang, Kim, Jung, Kim, Cho, Kim, Rhee,
  Park, and Park]{Lee2018}
Lee,~S.; Jang,~J.; Kim,~S.~I.; Jung,~S.~G.; Kim,~J.; Cho,~S.; Kim,~S.~W.;
  Rhee,~J.~Y.; Park,~K.~S.; Park,~T. Origin of Extremely Large
  Magnetoresistance in the Candidate Type-II Weyl Semimetal MoTe$_{2-x}$.
  \emph{Sci. Rep.} \textbf{2018}, \emph{8}, 13937, DOI:
  \doi{10.1038/s41598-018-32387-1}\relax
\mciteBstWouldAddEndPuncttrue
\mciteSetBstMidEndSepPunct{\mcitedefaultmidpunct}
{\mcitedefaultendpunct}{\mcitedefaultseppunct}\relax
\EndOfBibitem
\bibitem[Jindal \latin{et~al.}(2023)Jindal, Amartyajyoti, Zizhonh, Taniguchi,
  Watanabe, Hone, Birol, Fernandes, Dean, and Pasupathy]{Jindal2023}
Jindal,~A.; Amartyajyoti,~S.; Zizhonh,~L.; Taniguchi,~T.; Watanabe,~K.;
  Hone,~J.~C.; Birol,~T.; Fernandes,~R.~M.; Dean,~C.~R.; Pasupathy,~D.~A.,~A.
  N. and.~Rhodes Coupled ferroelectricity and superconductivity in bilayer
  T$_d$-MoTe$_2$. \emph{Nature} \textbf{2023}, 48--52, DOI:
  \doi{10.1038/s41586-022-05521-3}\relax
\mciteBstWouldAddEndPuncttrue
\mciteSetBstMidEndSepPunct{\mcitedefaultmidpunct}
{\mcitedefaultendpunct}{\mcitedefaultseppunct}\relax
\EndOfBibitem
\bibitem[Qi \latin{et~al.}(2016)Qi, Naumov, Ali, Rajamathi, Schnelle, Barkalov,
  Hanfland, Wu, Shekhar, Sun, Süß, Schmidt, Schwarz, Pippel, Werner,
  Hillebrand, Förster, Kampert, Parkin, Cava, \latin{et~al.} others]{Qi2016}
Qi,~Y. \latin{et~al.}  Superconductivity in Weyl Semimetal Candidate MoTe$_2$.
  \emph{Nat. Comms.} \textbf{2016}, \emph{7}, 11038, DOI:
  \doi{10.1038/ncomms11038}\relax
\mciteBstWouldAddEndPuncttrue
\mciteSetBstMidEndSepPunct{\mcitedefaultmidpunct}
{\mcitedefaultendpunct}{\mcitedefaultseppunct}\relax
\EndOfBibitem
\bibitem[Ruppert \latin{et~al.}(2014)Ruppert, Aslan, and Heinz]{Ruppert2014}
Ruppert,~C.; Aslan,~O.~B.; Heinz,~T.~F. Optical Properties and Band Gap of
  Single- and Few-Layer MoTe$_2$ Crystals. \emph{Nano Lett.} \textbf{2014},
  \emph{14}, 6231--6236, DOI: \doi{10.1021/nl502557g}\relax
\mciteBstWouldAddEndPuncttrue
\mciteSetBstMidEndSepPunct{\mcitedefaultmidpunct}
{\mcitedefaultendpunct}{\mcitedefaultseppunct}\relax
\EndOfBibitem
\bibitem[Lezama \latin{et~al.}(2015)Lezama, Arora, Ubaldini, Barreteau,
  Giannini, Potemski, and Morpurgo]{Lezama2015}
Lezama,~I.~G.; Arora,~A.; Ubaldini,~A.; Barreteau,~C.; Giannini,~E.;
  Potemski,~M.; Morpurgo,~A.~F. Indirect-to-Direct Band Gap Crossover in
  Few-Layer MoTe$_2$. \emph{Nano Lett.} \textbf{2015}, \emph{15}, 2336--2342,
  DOI: \doi{10.1021/nl5045007}\relax
\mciteBstWouldAddEndPuncttrue
\mciteSetBstMidEndSepPunct{\mcitedefaultmidpunct}
{\mcitedefaultendpunct}{\mcitedefaultseppunct}\relax
\EndOfBibitem
\bibitem[Mleczko \latin{et~al.}(2019)Mleczko, Yu, Smyth, Chen, Shin,
  Chatterjee, Tsai, Nishi, Wallace, and Pop]{Mleczko2019}
Mleczko,~M.~J.; Yu,~A.~C.; Smyth,~C.~M.; Chen,~V.; Shin,~Y.~C.; Chatterjee,~S.;
  Tsai,~Y.~C.; Nishi,~Y.; Wallace,~R.~M.; Pop,~E. Contact Engineering
  High-Performance n-Type MoTe$_2$ Transistors. \emph{Nano Lett.}
  \textbf{2019}, \emph{19}, 6352--6362, DOI:
  \doi{10.1021/acs.nanolett.9b02497}\relax
\mciteBstWouldAddEndPuncttrue
\mciteSetBstMidEndSepPunct{\mcitedefaultmidpunct}
{\mcitedefaultendpunct}{\mcitedefaultseppunct}\relax
\EndOfBibitem
\bibitem[Pradhan \latin{et~al.}(2014)Pradhan, Rhodes, Feng, Xin, Memaran, Moon,
  Terrones, Terrones, and Balicas]{Pradhan2014}
Pradhan,~N.~R.; Rhodes,~D.; Feng,~S.; Xin,~Y.; Memaran,~S.; Moon,~B.~H.;
  Terrones,~H.; Terrones,~M.; Balicas,~L. Field-Effect Transistors Based on
  Few-Layered $\alpha$-MoTe$_2$. \emph{ACS Nano} \textbf{2014}, \emph{8},
  5911--5920, DOI: \doi{10.1021/nn501013c}\relax
\mciteBstWouldAddEndPuncttrue
\mciteSetBstMidEndSepPunct{\mcitedefaultmidpunct}
{\mcitedefaultendpunct}{\mcitedefaultseppunct}\relax
\EndOfBibitem
\bibitem[Li \latin{et~al.}(2018)Li, Zhang, Xue, Zhou, and Yang]{Li2018}
Li,~N.; Zhang,~J.; Xue,~Y.; Zhou,~T.; Yang,~Z. Large Valley Polarization in
  Monolayer MoTe$_2$ on a Magnetic Substrate. \emph{Phys. Chem. Chem. Phys.}
  \textbf{2018}, \emph{20}, 3805--3812, DOI: \doi{10.1039/C7CP07610J}\relax
\mciteBstWouldAddEndPuncttrue
\mciteSetBstMidEndSepPunct{\mcitedefaultmidpunct}
{\mcitedefaultendpunct}{\mcitedefaultseppunct}\relax
\EndOfBibitem
\bibitem[Lin \latin{et~al.}(2014)Lin, Xu, Wang, Li, Yamamoto,
  Aparecido-Ferreira, Li, Sun, Nakaharai, Jian, Ueno, Tsukagoshi, Transistors,
  and in~Logic~Circuits]{Lin2014}
Lin,~Y.~F.; Xu,~Y.; Wang,~S.~T.; Li,~S.~L.; Yamamoto,~M.;
  Aparecido-Ferreira,~A.; Li,~W.; Sun,~H.; Nakaharai,~S.; Jian,~W.; Ueno,~B.;
  Tsukagoshi,~K.; Transistors,~K. A.~M.; in~Logic~Circuits,~T.~A. Ambipolar
  MoTe$_2$ Transistors and Their Applications in Logic Circuits. \emph{Adv.
  Mat.} \textbf{2014}, \emph{26}, 3263--3269, DOI:
  \doi{10.1002/adma.201305845}\relax
\mciteBstWouldAddEndPuncttrue
\mciteSetBstMidEndSepPunct{\mcitedefaultmidpunct}
{\mcitedefaultendpunct}{\mcitedefaultseppunct}\relax
\EndOfBibitem
\bibitem[Pace \latin{et~al.}(2021)Pace, Martini, Convertino, Keum, Forti,
  Pezzini, Fabbri, Miseikis, and Coletti]{Pace2021}
Pace,~S.; Martini,~L.; Convertino,~D.; Keum,~D.~H.; Forti,~S.; Pezzini,~S.;
  Fabbri,~F.; Miseikis,~V.; Coletti,~C. Synthesis of Large-Scale Monolayer
  1T'-MoTe$_2$ and Its Stabilization via Scalable hBN Encapsulation. \emph{ACS
  Nano} \textbf{2021}, \emph{15}, 4213--4225, DOI:
  \doi{10.1021/acsnano.0c05936}\relax
\mciteBstWouldAddEndPuncttrue
\mciteSetBstMidEndSepPunct{\mcitedefaultmidpunct}
{\mcitedefaultendpunct}{\mcitedefaultseppunct}\relax
\EndOfBibitem
\bibitem[Chi \latin{et~al.}(2019)Chi, Chen, Zhao, and Weng]{Chi2019}
Chi,~Z.; Chen,~H.; Zhao,~Q.; Weng,~Y.-X. Ultrafast carrier and phonon dynamics
  in few-layer 2H–MoTe$_2$. \emph{J. Chem. Phys.} \textbf{2019}, \emph{151},
  11470, DOI: \doi{10.1063/1.5115467}\relax
\mciteBstWouldAddEndPuncttrue
\mciteSetBstMidEndSepPunct{\mcitedefaultmidpunct}
{\mcitedefaultendpunct}{\mcitedefaultseppunct}\relax
\EndOfBibitem
\bibitem[Perlangeli \latin{et~al.}(2020)Perlangeli, Peli, Soranzio, Puntel,
  Parmigiani, and Cilento]{Perlangeli2020}
Perlangeli,~M.; Peli,~S.; Soranzio,~D.; Puntel,~D.; Parmigiani,~F.; Cilento,~F.
  Polarization-resolved broadband time-resolved optical spectroscopy for
  complex materials: application to the case of MoTe$_2$ polytypes. \emph{Opt.
  Exp.} \textbf{2020}, \emph{28}, 38541, DOI: \doi{10.1364/OE.385419}\relax
\mciteBstWouldAddEndPuncttrue
\mciteSetBstMidEndSepPunct{\mcitedefaultmidpunct}
{\mcitedefaultendpunct}{\mcitedefaultseppunct}\relax
\EndOfBibitem
\bibitem[Schulzetenberg and Johns(2021)Schulzetenberg, and
  Johns]{Schulzetenberg2021}
Schulzetenberg,~A.; Johns,~J. Chemical defects control the exciton lifetime in
  CVD grown, few-layer MoTe$_2$. \emph{Electron. Struct.} \textbf{2021},
  \emph{3}, 025001, DOI: \doi{10.1088/2516-1075/abaaf1}\relax
\mciteBstWouldAddEndPuncttrue
\mciteSetBstMidEndSepPunct{\mcitedefaultmidpunct}
{\mcitedefaultendpunct}{\mcitedefaultseppunct}\relax
\EndOfBibitem
\bibitem[Attar \latin{et~al.}(2020)Attar, Chang, Britz, Zhang, Lin,
  Krishnamoorthy, Linker, Fritz, Neumark, Kalia, Nakano, Ajayan, Vashishta,
  Bergmann, and Leone]{Attar2020}
Attar,~A.~R.; Chang,~H.-T.; Britz,~A.; Zhang,~X.; Lin,~M.-F.;
  Krishnamoorthy,~A.; Linker,~T.; Fritz,~D.; Neumark,~D.~M.; Kalia,~R.~K.;
  Nakano,~A.; Ajayan,~P.; Vashishta,~P.; Bergmann,~U.; Leone,~S.~R.
  Simultaneous Observation of Carrier-Specific Redistribution and Coherent
  Lattice Dynamics in 2H-MoTe$_2$ with Femtosecond Core-Level Spectroscopy.
  \emph{ACS Nano} \textbf{2020}, \emph{14}, 15829, DOI:
  \doi{10.1021/acsnano.0c06988}\relax
\mciteBstWouldAddEndPuncttrue
\mciteSetBstMidEndSepPunct{\mcitedefaultmidpunct}
{\mcitedefaultendpunct}{\mcitedefaultseppunct}\relax
\EndOfBibitem
\bibitem[Wang \latin{et~al.}(2021)Wang, Altmann, Gadermaier, Yang, Li,
  Ghirardini, Trovatello, Finazzi, Duò, Celebrano, Long, Akinwande, Prezhdo,
  Cerullo, and Dal~Conte]{Wang2021}
Wang,~Z.; Altmann,~P.; Gadermaier,~C.; Yang,~Y.; Li,~W.; Ghirardini,~L.;
  Trovatello,~C.; Finazzi,~M.; Duò,~L.; Celebrano,~M.; Long,~R.;
  Akinwande,~D.; Prezhdo,~O.~V.; Cerullo,~G.; Dal~Conte,~S. Phonon-Mediated
  Interlayer Charge Separation and Recombination in a MoSe$_2$/WSe$_2$
  Heterostructure. \emph{Nano Lett.} \textbf{2021}, \emph{21}, 2165--2173, DOI:
  \doi{10.1021/acs.nanolett.0c04955}\relax
\mciteBstWouldAddEndPuncttrue
\mciteSetBstMidEndSepPunct{\mcitedefaultmidpunct}
{\mcitedefaultendpunct}{\mcitedefaultseppunct}\relax
\EndOfBibitem
\bibitem[Froehlicher \latin{et~al.}(2015)Froehlicher, Lorchat, Fernique, Joshi,
  Molina-Sánchez, Wirtz, and Berciaud]{Froehlicher2015}
Froehlicher,~G.; Lorchat,~E.; Fernique,~F.; Joshi,~C.; Molina-Sánchez,~A.;
  Wirtz,~L.; Berciaud,~S. Unified Description of the Optical Phonon Modes in
  N-Layer MoTe$_2$. \emph{Nano Lett.} \textbf{2015}, \emph{15}, 6481, DOI:
  \doi{10.1021/acs.nanolett.5b02683}\relax
\mciteBstWouldAddEndPuncttrue
\mciteSetBstMidEndSepPunct{\mcitedefaultmidpunct}
{\mcitedefaultendpunct}{\mcitedefaultseppunct}\relax
\EndOfBibitem
\bibitem[Wilson and Yoffe(1969)Wilson, and Yoffe]{Wilson1969}
Wilson,~J.~A.; Yoffe,~A.~D. The transition metal dichalcogenides discussion and
  interpretation of the observed optical, electrical and structural properties.
  \emph{Advances in Physics} \textbf{1969}, \emph{18}, 193--335, DOI:
  \doi{10.1080/00018736900101307}\relax
\mciteBstWouldAddEndPuncttrue
\mciteSetBstMidEndSepPunct{\mcitedefaultmidpunct}
{\mcitedefaultendpunct}{\mcitedefaultseppunct}\relax
\EndOfBibitem
\bibitem[Pogna \latin{et~al.}(2016)Pogna, Marsili, De~Fazio, Dal~Conte,
  Manzoni, Sangalli, Yoon, Lombardo, Ferrari, Marini, Cerullo, and
  Prezzi]{Pogna2016}
Pogna,~E. A.~A.; Marsili,~M.; De~Fazio,~D.; Dal~Conte,~S.; Manzoni,~C.;
  Sangalli,~D.; Yoon,~D.; Lombardo,~A.; Ferrari,~A.~C.; Marini,~A.;
  Cerullo,~G.; Prezzi,~D. Photo-Induced Bandgap Renormalization Governs the
  Ultrafast Response of Single-Layer MoS$_2$. \emph{ACS Nano} \textbf{2016},
  \emph{10}, 1182--1188, DOI: \doi{10.1021/acsnano.5b06488}\relax
\mciteBstWouldAddEndPuncttrue
\mciteSetBstMidEndSepPunct{\mcitedefaultmidpunct}
{\mcitedefaultendpunct}{\mcitedefaultseppunct}\relax
\EndOfBibitem
\bibitem[Trovatello \latin{et~al.}(2022)Trovatello, Katsch, Li, Zhu, Knorr,
  Cerullo, and Dal~Conte]{Trovatello2022}
Trovatello,~C.; Katsch,~F.; Li,~Q.; Zhu,~X.; Knorr,~A.; Cerullo,~G.;
  Dal~Conte,~S. Disentangling Many-Body Effects in the Coherent Optical
  Response of 2D Semiconductors. \emph{Nano Letters} \textbf{2022}, \emph{22},
  5322--5329, DOI: \doi{10.1021/acs.nanolett.2c01309}\relax
\mciteBstWouldAddEndPuncttrue
\mciteSetBstMidEndSepPunct{\mcitedefaultmidpunct}
{\mcitedefaultendpunct}{\mcitedefaultseppunct}\relax
\EndOfBibitem
\bibitem[Hase \latin{et~al.}(2000)Hase, Ishioka, Kitajima, Ushida, and
  Hishita]{Hase2000}
Hase,~M.; Ishioka,~K.; Kitajima,~M.; Ushida,~K.; Hishita,~S. Dephasing of
  Coherent Phonons by Lattice Defects in Bismuth Films. \emph{Appl. Phys.
  Lett.} \textbf{2000}, \emph{76}, 1258, DOI: \doi{10.1063/1.126002}\relax
\mciteBstWouldAddEndPuncttrue
\mciteSetBstMidEndSepPunct{\mcitedefaultmidpunct}
{\mcitedefaultendpunct}{\mcitedefaultseppunct}\relax
\EndOfBibitem
\bibitem[Dhar \latin{et~al.}(1994)Dhar, Rogers, and Nelson]{Dhar1994}
Dhar,~L.; Rogers,~J.~A.; Nelson,~K.~A. Time-Resolved Vibrational Spectroscopy
  in the Impulsive Limit. \emph{Chem. Rev.} \textbf{1994}, \emph{94}, 157--193,
  DOI: \doi{10.1021/cr00025a006}\relax
\mciteBstWouldAddEndPuncttrue
\mciteSetBstMidEndSepPunct{\mcitedefaultmidpunct}
{\mcitedefaultendpunct}{\mcitedefaultseppunct}\relax
\EndOfBibitem
\bibitem[Zeiger \latin{et~al.}(1992)Zeiger, Vidal, Cheng, Ippen, Dresselhaus,
  and Dresselhaus]{Zeiger1992}
Zeiger,~H.~J.; Vidal,~J.; Cheng,~T.~K.; Ippen,~E.~P.; Dresselhaus,~G.;
  Dresselhaus,~M.~S. Theory for Displacive Excitation of Coherent Phonons.
  \emph{Phys. Rev. B} \textbf{1992}, \emph{45}, 768--778, DOI:
  \doi{10.1103/PhysRevB.45.768}\relax
\mciteBstWouldAddEndPuncttrue
\mciteSetBstMidEndSepPunct{\mcitedefaultmidpunct}
{\mcitedefaultendpunct}{\mcitedefaultseppunct}\relax
\EndOfBibitem
\bibitem[Molina-Sánchez \latin{et~al.}(2020)Molina-Sánchez, Catarina,
  Sangalli, and Fernández-Rossier]{Molina-Sanchez2020}
Molina-Sánchez,~A.; Catarina,~G.; Sangalli,~D.; Fernández-Rossier,~J.
  Magneto-optical response of chromium trihalide monolayers: chemical trends.
  \emph{J. Mater. Chem. C} \textbf{2020}, \emph{8}, 8856--8863, DOI:
  \doi{10.1039/D0TC01322F}\relax
\mciteBstWouldAddEndPuncttrue
\mciteSetBstMidEndSepPunct{\mcitedefaultmidpunct}
{\mcitedefaultendpunct}{\mcitedefaultseppunct}\relax
\EndOfBibitem
\bibitem[Bernardi \latin{et~al.}(2013)Bernardi, Palummo, and
  Grossman]{Benardi2013}
Bernardi,~M.; Palummo,~M.; Grossman,~J.~C. Extraordinary Sunlight Absorption
  and One Nanometer Thick Photovoltaics Using Two-Dimensional Monolayer
  Materials. \emph{Nano Lett.} \textbf{2013}, \emph{13}, 3664--3670, DOI:
  \doi{10.1021/nl401544y}\relax
\mciteBstWouldAddEndPuncttrue
\mciteSetBstMidEndSepPunct{\mcitedefaultmidpunct}
{\mcitedefaultendpunct}{\mcitedefaultseppunct}\relax
\EndOfBibitem
\bibitem[Trovatello \latin{et~al.}(2020)Trovatello, Katsch, Borys, Selig, Yao,
  Borrego-Varillas, Scotognella, Kriegel, Yan, Zettl, Schuck, Knorr, Cerullo,
  and Conte]{Trovatello2020b}
Trovatello,~C.; Katsch,~F.; Borys,~N.~J.; Selig,~M.; Yao,~K.;
  Borrego-Varillas,~R.; Scotognella,~F.; Kriegel,~I.; Yan,~A.; Zettl,~A.;
  Schuck,~P.~J.; Knorr,~A.; Cerullo,~G.; Conte,~S.~D. The ultrafast onset of
  exciton formation in 2D semiconductors. \emph{Nat. Comm.} \textbf{2020},
  \emph{11}, 5277, DOI: \doi{10.1038/s41467-020-18835-5}\relax
\mciteBstWouldAddEndPuncttrue
\mciteSetBstMidEndSepPunct{\mcitedefaultmidpunct}
{\mcitedefaultendpunct}{\mcitedefaultseppunct}\relax
\EndOfBibitem
\bibitem[Hohenberg and Kohn(1964)Hohenberg, and Kohn]{HohenbergKohn1964}
Hohenberg,~P.; Kohn,~W. Inhomogeneous Electron Gas. \emph{Phys. Rev.}
  \textbf{1964}, \emph{136}, B864--B871, DOI:
  \doi{10.1103/PhysRev.136.B864}\relax
\mciteBstWouldAddEndPuncttrue
\mciteSetBstMidEndSepPunct{\mcitedefaultmidpunct}
{\mcitedefaultendpunct}{\mcitedefaultseppunct}\relax
\EndOfBibitem
\bibitem[Kohn and Sham(1965)Kohn, and Sham]{KohnSham1965}
Kohn,~W.; Sham,~L.~J. Self-Consistent Equations Including Exchange and
  Correlation Effects. \emph{Phys. Rev.} \textbf{1965}, \emph{140},
  A1133--A1138, DOI: \doi{10.1103/PhysRev.140.A1133}\relax
\mciteBstWouldAddEndPuncttrue
\mciteSetBstMidEndSepPunct{\mcitedefaultmidpunct}
{\mcitedefaultendpunct}{\mcitedefaultseppunct}\relax
\EndOfBibitem
\bibitem[Baroni \latin{et~al.}(2001)Baroni, de~Gironcoli, Dal~Corso, and
  Giannozzi]{Baroni2001}
Baroni,~S.; de~Gironcoli,~S.; Dal~Corso,~A.; Giannozzi,~P. Phonons and related
  crystal properties from density-functional perturbation theory. \emph{Rev.
  Mod. Phys.} \textbf{2001}, \emph{73}, 515--562, DOI:
  \doi{10.1103/RevModPhys.73.515}\relax
\mciteBstWouldAddEndPuncttrue
\mciteSetBstMidEndSepPunct{\mcitedefaultmidpunct}
{\mcitedefaultendpunct}{\mcitedefaultseppunct}\relax
\EndOfBibitem
\bibitem[Giannozzi \latin{et~al.}(2009)Giannozzi, Baroni, Bonini, Calandra,
  Car, Cavazzoni, Ceresoli, Chiarotti, Cococcioni, Dabo, Corso, de~Gironcoli,
  Fabris, Fratesi, Gebauer, Gerstmann, Gougoussis, Kokalj, Lazzeri,
  Martin-Samos, Marzari, Mauri, Mazzarello, Paolini, Pasquarello, Paulatto,
  Sbraccia, Scandolo, Sclauzero, Seitsonen, Smogunov, Umari, and
  Wentzcovitch]{Giannozzi2009}
Giannozzi,~P. \latin{et~al.}  QUANTUM ESPRESSO: a modular and open-source
  software project for quantum simulations of materials. \emph{J. Phys:
  Condensed Matter} \textbf{2009}, \emph{21}, 395502, DOI:
  \doi{10.1088/0953-8984/21/39/395502}\relax
\mciteBstWouldAddEndPuncttrue
\mciteSetBstMidEndSepPunct{\mcitedefaultmidpunct}
{\mcitedefaultendpunct}{\mcitedefaultseppunct}\relax
\EndOfBibitem
\bibitem[Giannozzi \latin{et~al.}(2017)Giannozzi, Andreussi, Brumme, Bunau,
  Nardelli, Calandra, Car, Cavazzoni, Ceresoli, Cococcioni, Colonna, Carnimeo,
  Corso, de~Gironcoli, Delugas, DiStasio, Ferretti, Floris, Fratesi, Fugallo,
  Gebauer, Gerstmann, Giustino, Gorni, Jia, Kawamura, Ko, Kokalj,
  Kü{\c{c}}ükbenli, Lazzeri, Marsili, Marzari, Mauri, Nguyen, Nguyen, de-la
  Roza, Paulatto, Ponc{\'{e}}, Rocca, Sabatini, Santra, Schlipf, Seitsonen,
  Smogunov, Timrov, Thonhauser, Umari, Vast, Wu, and Baroni]{Giannozzi2017}
Giannozzi,~P. \latin{et~al.}  Advanced capabilities for materials modelling
  with Quantum {ESPRESSO}. \emph{J. Phys.: Condensed Matter} \textbf{2017},
  \emph{29}, 465901, DOI: \doi{10.1088/1361-648x/aa8f79}\relax
\mciteBstWouldAddEndPuncttrue
\mciteSetBstMidEndSepPunct{\mcitedefaultmidpunct}
{\mcitedefaultendpunct}{\mcitedefaultseppunct}\relax
\EndOfBibitem
\bibitem[Giannozzi \latin{et~al.}(2020)Giannozzi, Baseggio, Bonfà, Brunato,
  Car, Carnimeo, Cavazzoni, de~Gironcoli, Delugas, Ferrari~Ruffino, Ferretti,
  Marzari, Timrov, Urru, and Baroni]{Giannozzi2020}
Giannozzi,~P.; Baseggio,~O.; Bonfà,~P.; Brunato,~D.; Car,~R.; Carnimeo,~I.;
  Cavazzoni,~C.; de~Gironcoli,~S.; Delugas,~P.; Ferrari~Ruffino,~F.;
  Ferretti,~A.; Marzari,~N.; Timrov,~I.; Urru,~A.; Baroni,~S. Quantum ESPRESSO
  toward the exascale. \emph{J. Chem. Phys.} \textbf{2020}, \emph{152}, 154105,
  DOI: \doi{10.1063/5.0005082}\relax
\mciteBstWouldAddEndPuncttrue
\mciteSetBstMidEndSepPunct{\mcitedefaultmidpunct}
{\mcitedefaultendpunct}{\mcitedefaultseppunct}\relax
\EndOfBibitem
\bibitem[Perdew \latin{et~al.}(1996)Perdew, Burke, and Ernzerhof]{Perdew1996}
Perdew,~J.~P.; Burke,~K.; Ernzerhof,~M. Generalized Gradient Approximation Made
  Simple. \emph{Phys. Rev. Lett.} \textbf{1996}, \emph{77}, 3865--3868, DOI:
  \doi{10.1103/PhysRevLett.77.3865}\relax
\mciteBstWouldAddEndPuncttrue
\mciteSetBstMidEndSepPunct{\mcitedefaultmidpunct}
{\mcitedefaultendpunct}{\mcitedefaultseppunct}\relax
\EndOfBibitem
\bibitem[{van Setten} \latin{et~al.}(2018){van Setten}, Giantomassi, Bousquet,
  Verstraete, Hamann, Gonze, and Rignanese]{vanSetten2018}
{van Setten},~M.; Giantomassi,~M.; Bousquet,~E.; Verstraete,~M.; Hamann,~D.;
  Gonze,~X.; Rignanese,~G.-M. The PseudoDojo: Training and grading a 85 element
  optimized norm-conserving pseudopotential table. \emph{Comp. Phys. Comms.}
  \textbf{2018}, \emph{226}, 39--54, DOI: \doi{10.1016/j.cpc.2018.01.012}\relax
\mciteBstWouldAddEndPuncttrue
\mciteSetBstMidEndSepPunct{\mcitedefaultmidpunct}
{\mcitedefaultendpunct}{\mcitedefaultseppunct}\relax
\EndOfBibitem
\bibitem[Onida \latin{et~al.}(2002)Onida, Reining, and Rubio]{Onida2002}
Onida,~G.; Reining,~L.; Rubio,~A. Electronic excitations: density-functional
  versus many-body Green's-function approaches. \emph{Rev. Mod. Phys.}
  \textbf{2002}, \emph{74}, 601--659, DOI:
  \doi{10.1103/RevModPhys.74.601}\relax
\mciteBstWouldAddEndPuncttrue
\mciteSetBstMidEndSepPunct{\mcitedefaultmidpunct}
{\mcitedefaultendpunct}{\mcitedefaultseppunct}\relax
\EndOfBibitem
\bibitem[Reining(2018)]{Reining2018}
Reining,~L. The GW approximation: content, successes and limitations.
  \emph{WIREs Comput. Mol. Sci.} \textbf{2018}, \emph{8}, e1344, DOI:
  \doi{10.1002/wcms.1344}\relax
\mciteBstWouldAddEndPuncttrue
\mciteSetBstMidEndSepPunct{\mcitedefaultmidpunct}
{\mcitedefaultendpunct}{\mcitedefaultseppunct}\relax
\EndOfBibitem
\bibitem[Marsili \latin{et~al.}(2021)Marsili, Molina-S\'anchez, Palummo,
  Sangalli, and Marini]{Marsili2021}
Marsili,~M.; Molina-S\'anchez,~A.; Palummo,~M.; Sangalli,~D.; Marini,~A.
  Spinorial formulation of the $GW$-BSE equations and spin properties of
  excitons in two-dimensional transition metal dichalcogenides. \emph{Phys.
  Rev. B} \textbf{2021}, \emph{103}, 155152, DOI:
  \doi{10.1103/PhysRevB.103.155152}\relax
\mciteBstWouldAddEndPuncttrue
\mciteSetBstMidEndSepPunct{\mcitedefaultmidpunct}
{\mcitedefaultendpunct}{\mcitedefaultseppunct}\relax
\EndOfBibitem
\bibitem[Miranda \latin{et~al.}(2017)Miranda, Reichardt, Froehlicher,
  Molina-Sánchez, Berciaud, and Wirtz]{Miranda2017}
Miranda,~H. P.~C.; Reichardt,~S.; Froehlicher,~G.; Molina-Sánchez,~A.;
  Berciaud,~S.; Wirtz,~L. Quantum Interference Effects in Resonant Raman
  Spectroscopy of Single- and Triple-Layer MoTe$_2$ from First-Principles.
  \emph{Nano Lett.} \textbf{2017}, \emph{17}, 2381--2388, DOI:
  \doi{10.1021/acs.nanolett.6b05345}\relax
\mciteBstWouldAddEndPuncttrue
\mciteSetBstMidEndSepPunct{\mcitedefaultmidpunct}
{\mcitedefaultendpunct}{\mcitedefaultseppunct}\relax
\EndOfBibitem
\bibitem[Sangalli \latin{et~al.}(2019)Sangalli, Ferretti, Miranda, Attaccalite,
  Marri, Cannuccia, Melo, Marsili, Paleari, Marrazzo, Prandini, Bonfà, Atambo,
  Affinito, Palummo, Molina-Sánchez, Hogan, Grüning, Varsano, and
  Marini]{Sangalli2019}
Sangalli,~D. \latin{et~al.}  Many-body perturbation theory calculations using
  the yambo code. \emph{J. Phys: Condensed Matter} \textbf{2019}, \emph{31},
  325902, DOI: \doi{10.1088/1361-648X/ab15d0}\relax
\mciteBstWouldAddEndPuncttrue
\mciteSetBstMidEndSepPunct{\mcitedefaultmidpunct}
{\mcitedefaultendpunct}{\mcitedefaultseppunct}\relax
\EndOfBibitem
\bibitem[Monkhorst and Pack(1976)Monkhorst, and Pack]{Monkhorst1976}
Monkhorst,~H.~J.; Pack,~J.~D. Special points for Brillouin-zone integrations.
  \emph{Phys. Rev. B} \textbf{1976}, \emph{13}, 5188--5192, DOI:
  \doi{10.1103/PhysRevB.13.5188}\relax
\mciteBstWouldAddEndPuncttrue
\mciteSetBstMidEndSepPunct{\mcitedefaultmidpunct}
{\mcitedefaultendpunct}{\mcitedefaultseppunct}\relax
\EndOfBibitem
\end{mcitethebibliography}




\section*{Supplementary material}

# Supporting Information for:

# Strong Coupling of Coherent Phonons to Excitons in Semiconducting Monolayer MoTe$_2$.

Charles J. Sayers,*,† Armando Genco,† Chiara Trovatello,†,∥ Stefano Dal Conte,† Vladislav Khaustov,‡,⊥ Jorge Cervantes-Villanueva,¶ Davide Sangalli,§ Alejandro Molina-Sanchez,¶ Camilla Coletti,‡,# Christoph Gadermaier,† and Giulio Cerullo†

†*Dipartimento di Fisica, Politecnico di Milano, 20133 Milano, Italy*
‡*Center for Nanotechnology Innovation @ NEST, Istituto Italiano di Tecnologia, 56127 Pisa, Italy*
¶*Institute of Materials Science (ICMUV), University of Valencia, Catedrático Beltrán 2, E-46980 Valencia, Spain*
§*Istituto di Struttura della Materia-CNR (ISM-CNR), Division of Ultrafast Processes in Materials (FLASHit), Area della Ricerca di Roma 1, Monterotondo Scalo, Italy*
∥*Department of Mechanical Engineering, Columbia University, New York, New York 10027, United States*
⊥*Scuola Normale Superiore, Piazza San Silvestro 12, 56127 Pisa, Italy*
#*Graphene Labs, Istituto Italiano di Tecnologia, 16163 Genova, Italy*

E-mail: charles.sayers@polimi.it

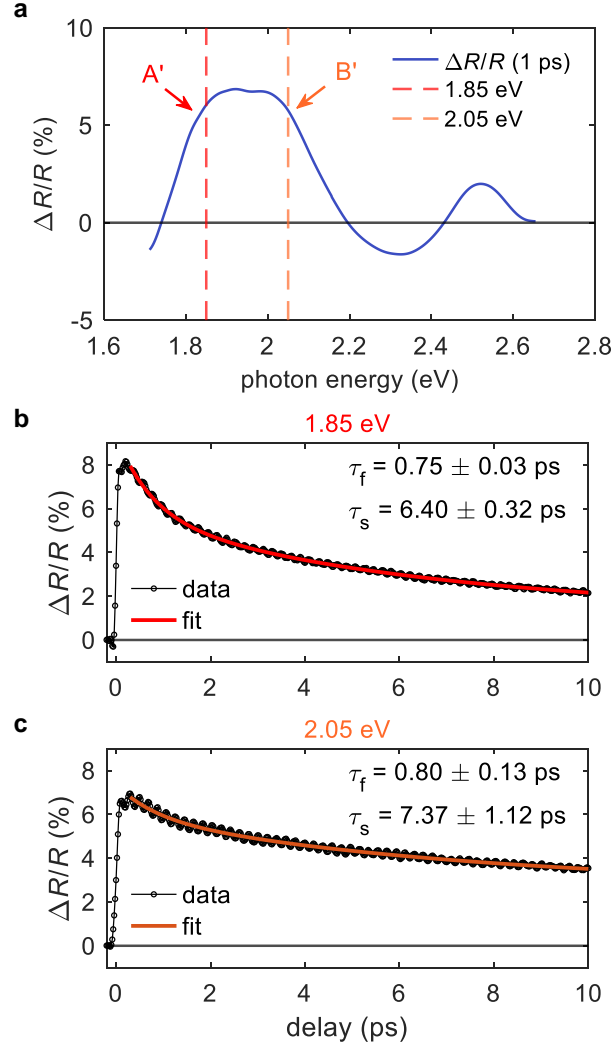

Figure S1: **Determination of the exciton dynamics by multi-exponential fit.** (**a**) Transient $\Delta R/R$ spectrum at 1 ps. Vertical dashed lines indicate photon energies 1.85 and 2.05 eV selected for multi-exponential fitting which correspond to the approximate positions of the photobleaching (PB) signal of the A' and B' excitonic resonances. (**b**) - (**c**) Fit of the PB signal for each photon energy using a bi-exponential decay function $A_\text{f} e^{-t/\tau_\text{f}} + A_\text{s} e^{-t/\tau_\text{s}} + y_0$ that consists of a fast ($\tau_\text{f}$) and slow ($\tau_\text{s}$) component and a offset value, $y_0$. Data for $t > 0.3$ ps was used to fit only the decay dynamics. The results of the fitting, including associated errors, are reported in the figure panels.

2

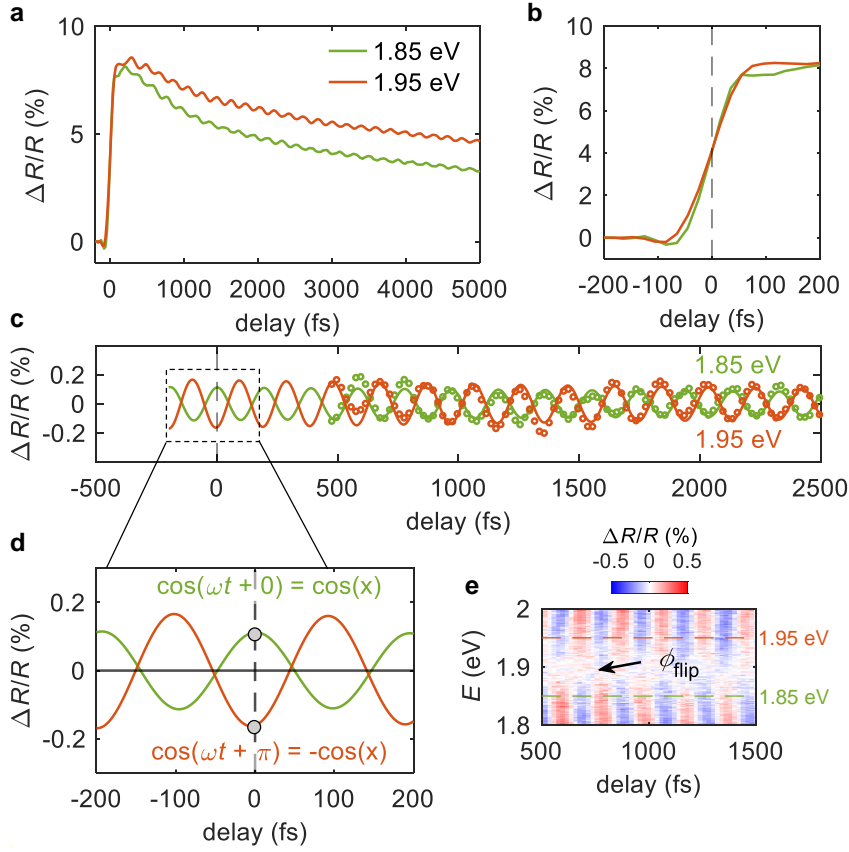

Figure S2: **Determination of the oscillatory phase.** (**a**) Dynamics at suitable probe photon energies of 1.85 eV and 1.95 eV are selected for the low and high energy side of the phase flip which occurs at $\sim 1.9$ eV, as reported in the main text [see also panel (e)]. (**b**) The zero delay ($t = 0$) is confirmed by the half-rise time of the incoherent $\Delta R/R$ signal as expected for a Gaussian pulse width. (**c**) The coherent component dynamics at the selected photon energies (open circles) are fitted with a damped sinusoidal function (solid lines) in the range $+500$ to $+5000$ fs and then extrapolated to negative delay. (**d**) Zoom of the extrapolated fit showing a maximum or a minimum in the oscillatory signal at $t = 0$ for the 1.85 eV and 1.95 eV traces respectively, confirming the cosine nature of the coherent oscillations, and a $\pi$ phase shift across $\sim 1.9$ eV. (**e**) Portion of the coherent component map as reported in the main text close to the phase flip energy at $\sim 1.9$ eV, highlighting the photon energies selected for the phase analysis.

3

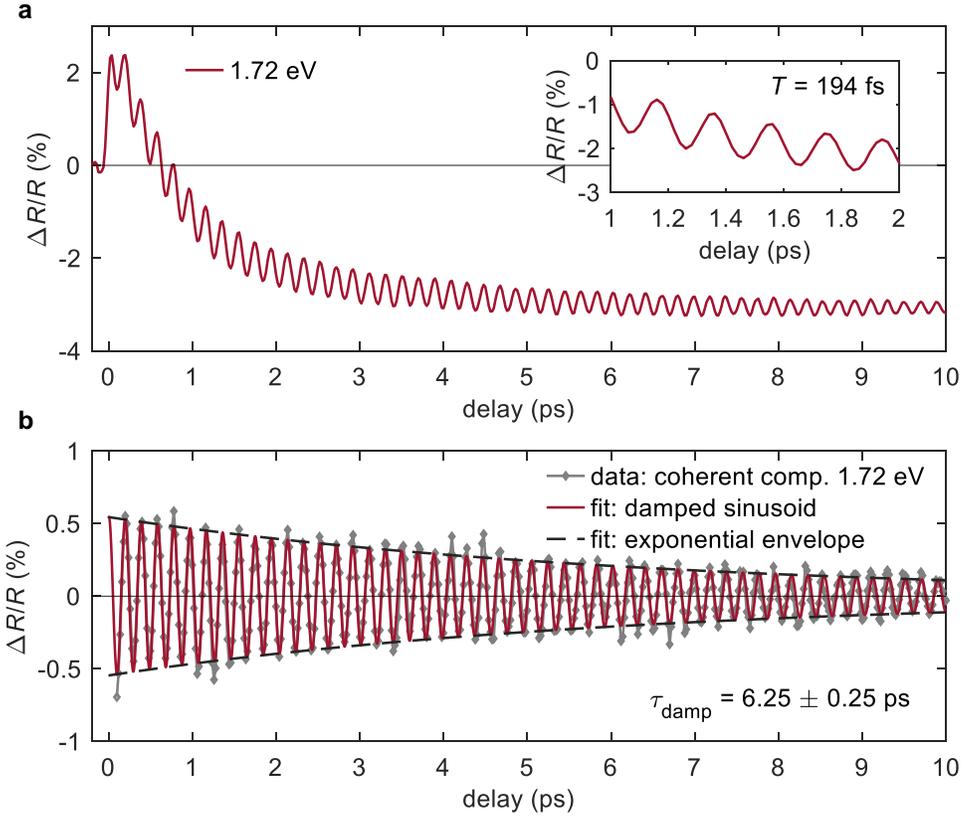

Figure S3: **Determination of the oscillatory damping time.** (**a**) Dynamics selected at 1.72 eV near the maximum oscillatory signal, as seen in the main text. The inset shows a zoom of the data highlighting the $T = 194$ fs period (5.15 THz frequency). (**b**) Coherent component (grey dots) of the $\Delta R/R$ signal after subtraction of an exponential fit to the incoherent component. The coherent component data is fitted with a damped sinusoidal function $y(t) = Ae^{-t/\tau_{\text{damp}}} \cos(\omega t + \phi)$ (solid red line) which gives a damping time of $\tau_{damp} = 6.25 \pm 0.25$ ps. The exponential envelope is shown for completeness (dashed lines).

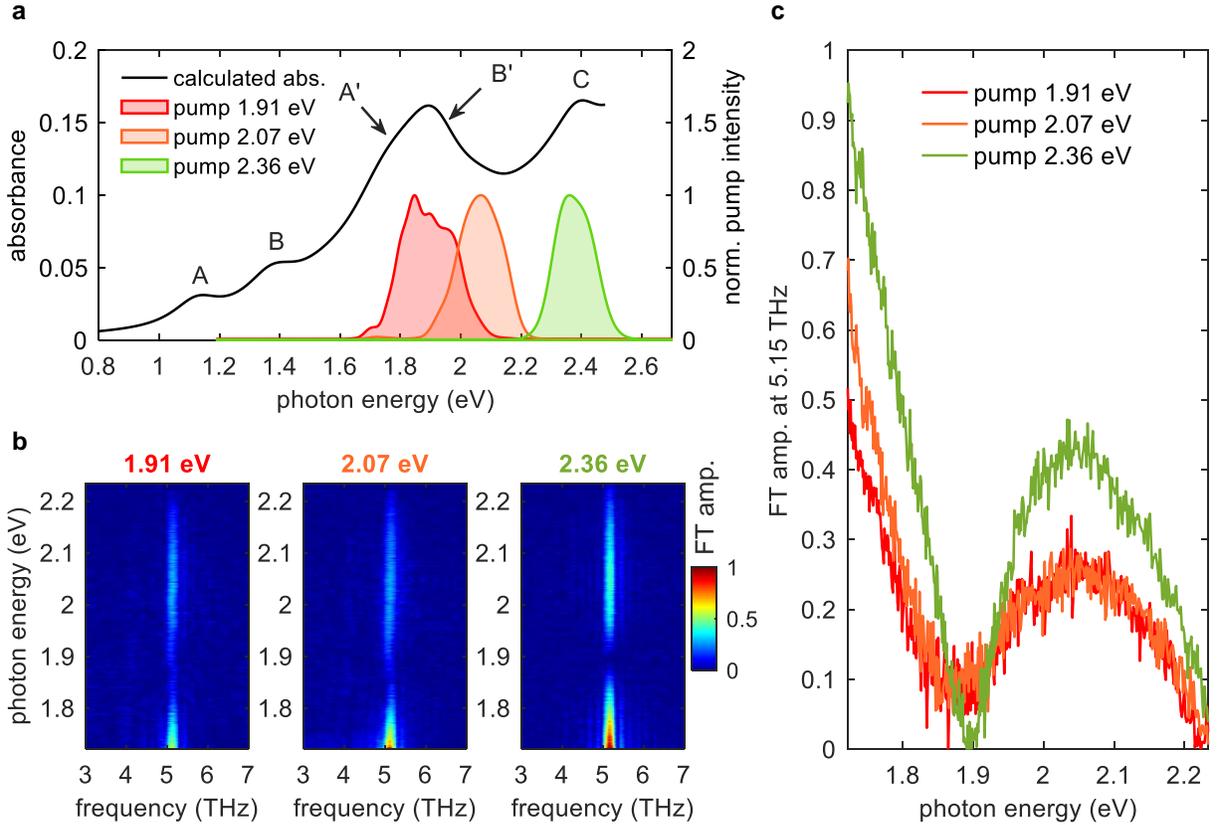

Figure S4: **Excitation energy dependence of the oscillation amplitude.** (**a**) Calculated optical absorbance (left axis) for the equilibrium structure of $2H$-MoTe$_2$. Normalized broadband pump spectra (right axis) with centre energies indicated in the legend. (**b**) Fourier transform (FT) map of the oscillatory signal component for each excitation (pump) energy. The pump fluence was $\sim 500\,\mu\text{J}\,\text{cm}^{-2}$ in all cases. (**c**) FT amplitude at 5.15 THz extracted from panel b for each excitation energy, revealing a similar spectral profile in all cases, and a slightly enhanced amplitude for higher pump energies.



\end{document}